\documentclass[12pt]{article}
\usepackage{psfrag,amsmath,amssymb,enumerate,epsfig}
\makeatletter
\renewcommand{\section}{\setcounter{equation}{0}\@startsection
  {section}%
  {1}%
  {0pt}%
  {-1\baselineskip}%
  {0.4\baselineskip}%
  {\bfseries\large}}%
\renewcommand{\subsection}{\@startsection
  {subsection}%
  {2}%
  {0pt}%
  {-0.75\baselineskip}%
  {0.2\baselineskip}%
  {\bfseries}}%
\renewcommand{\subsubsection}{\@startsection
  {subsubsection}%
  {3}%
  {0pt}%
  {-0.5\baselineskip}%
  {0.1\baselineskip}%
  {\sc}}%
\makeatother

\makeatletter
 \newcommand\figcaption{\def\@captype{figure}\caption}
\makeatother

\def\ak{{\mathfrak a}}
\def\fk{{\mathfrak f}}
\def\lk{{\mathfrak l}}

\def\a{\alpha}
\def\b{\beta}
\def\d{\delta}

\def\ga{\gamma}

\def\la{\lambda}

\def\m{\mu}
\def\n{\nu}
\def\r{\rho}
\def\s{\sigma}

\def\th{\theta}

\def\Dirac{{D\mkern-12mu/}}

\def\pslash{{p\mkern-8mu/}{\!}}

\def\pslash  {{p\mkern-7mu/}}

\def\id{{\rm{I}\!\rm{I}}}

\def\id3x{\int\!\! d^3\!\vec{x}}
\def\idx{\int\!\! d^4\!x}
\def\iDx{\int\!\! d^D\!x}

\def\RR{{\rm I\!\!\, R}}

\def\unit{{\rm 1\hskip-3pt I}}

\def\rig>{\right>}

\newcommand{\bea}{\begin{eqnarray}}
\newcommand{\eea}{\end{eqnarray}}
\newcommand{\beann}{\begin{eqnarray*}}
\newcommand{\eeann}{\end{eqnarray*}}
\newcommand{\ba}{\begin{array}}
\newcommand{\ea}{\end{array}}

\newcommand{\Tr}{\mathbf{Tr}}

\def\g5{\gamma_{5}}

\def\pslash  {{p\mkern-7mu/}}

\def\RR{{\rm I}\!{\rm R}}

\def\idx3{\int\! d^{3}\!\vec{x}\,}
\def\idx{\int\! d^{4}\!x\,}

 \def\Dirac{{D\mkern-12mu/}\,}

 \def\pslash  {{p\mkern-7mu/}}

 \def\RR{{\rm I}\!{\rm R}}

 \def\Da {{\partial}_{\alpha}}
 \def\Db {{\partial}_{\beta}}
 \def\Dm {{\partial}_{\mu}}
 
 \def\g {\gamma}

 \def\mi {{\mu_1}}
\def\mii {{\mu_2}}
 \def\miii {{\mu_3}}

 \def\a {\alpha}
\def\b {\beta}
\def\r {\rho}
 \def\s {\sigma}

\def\RR{{\rm I}\!{\rm R}}

 \def\Tr{\text{Tr}}

\begin{document}

\begin{titlepage}
\hfill{NSF-KITP-09-112}\\
\rightline{FTI/UCM 90-2009}\vglue 10pt
\begin{center}

{\Large \bf Noncommutative ${\cal N}=1$ super Yang-Mills,
 the Seiberg-Witten map and UV divergences.}\\
\vskip 0.5 true cm {\rm C.P. Mart\'{\i}n$^{\dagger,}$\footnote{E-mail:carmelo@elbereth.fis.ucm.es}
and  C. Tamarit$^{\dagger\dagger,}$}\footnote{E-mail: tamarit@kitp.ucsb.edu}
\vskip 0.3 true cm $^\dagger${\it Departamento de F\'{\i}sica Te\'orica I,
Facultad de Ciencias F\'{\i}sicas\\
Universidad Complutense de Madrid,
 28040 Madrid, Spain}
 \vskip 0.3 true cm $^{\dagger\dagger}${\it Kavli Institute for Theoretical Physics, University of California\\
 Santa Barbara, CA, 93106-4030, USA}\\
\vskip 0.5 true cm

{\leftskip=50pt \rightskip=50pt \noindent Classically, the  dual under the Seiberg-Witten map of noncommutative $U(N),$ ${\cal N}=1$ super Yang-Mills theory is a field theory with ordinary gauge symmetry whose  fields carry, however, a $\theta$-deformed nonlinear realisation of the ${\cal N}=1$ supersymmetry algebra in four dimensions. For the latter theory  we work out at one-loop and first order in the noncommutative parameter matrix $\theta^{\mu\nu}$ the UV divergent part of its effective action in the
background-field gauge, and, for  $N\neq1$, we show that for finite values of $N$ the gauge sector fails to be renormalisable; however, in the large $N$ limit the full theory is renormalisable, in keeping with the expectations raised by the quantum behaviour of the theory's noncommutative classical dual. We also 
obtain --for $N\geq3$, the case with $N=2$ being trivial-- the UV divergent part of the effective action of the $SU(N)$  noncommutative theory in the enveloping-algebra formalism that is obtained from the previous  ordinary $U(N)$ theory by removing the $U(1)$ degrees of freedom. This noncommutative $SU(N)$ theory is also renormalisable.

\par }
\end{center}
\vspace{5pt} \noindent
\end{titlepage}


\setcounter{page}{2}
\section{Introduction}

    Noncommutative gauge theories are known to arise as low energy limits of (super)string theory \cite{Seiberg:1999vs,Chu:1999ij}, and they are interesting on their own as  examples of nonlocal theories. One of their intriguing features is that  noncommutative U(N) gauge theories, considered as effective descriptions of the dynamics of D-branes with Neveu-Schwarz backgrounds, are known to have a dual description in terms of fields with ordinary gauge invariance \cite{Seiberg:1999vs}. This equivalence, which can be traced back to the possibility of choosing different yet equivalent regularisations of the D-Brane effective action, can be formulated by means of a map which relates noncommutative and ordinary gauge fields in a way consistent with their respective gauge symmetries, so that orbits of noncommutative gauge transformations are mapped into  orbits of ordinary gauge transformations. These maps are called Seiberg-Witten maps. Their role linking different DBI actions has also been shown to hold, at least to a certain approximation, in the ${\cal N}=1$ supersymmetric case \cite{Martin:2008xa}. In principle, this equivalence holds for the D-Brane effective actions, but one may wonder whether it also holds, at the quantum level, for the noncommutative gauge theories that do not involve the higher order terms present in the DBI actions.

	The idea of mapping noncommutative to ordinary gauge symmetries was the starting point for the formulation of noncommutative gauge theories for arbitrary gauge groups by means of Seiberg-Witten maps pioneered in refs.~\cite{Madore:2000en,Jurco:2000ja,Jurco:2001rq}. In the ``standard'' formalism, closure under gauge transformations restricts the gauge groups to be U(N) and the representations to be (anti-)fundamental or bi-(anti)-fundamental, while the formalism which makes use of Seiberg-Witten maps, also referred to as the enveloping algebra formalism, makes it possible to consider arbitrary gauge groups and representations by mapping the enveloping-algebra valued noncommutative gauge fields to ordinary Lie-algebra valued gauge fields.
	
   	The  quantum properties of noncommutative gauge theories, both in the standard and enveloping algebra approaches, have been analysed in many works. Concerning the standard approach, nonsupersymmetric noncommutative U(N) Yang-Mills theories are plagued by pathological IR divergences coming from the UV/IR mixing effect \cite{Minwalla:1999px}, which are suppressed in the large N limit, in which only planar diagrams contribute and the sole effect of noncommutativity is  producing phase factors  depending on the external momenta which can be taken out of the loop integrals. Noncommutative supersymmetric gauge theories \cite{Nishimura:2003jd} exhibit a better behaviour in the infrared, as the problematic divergences are milder or altogether absent \cite{Matusis:2000jf,Ruiz:2000hu, Zanon:2000nq}.
These milder noncommutative IR divergences are logarithmic and can be integrated leading to a consistent renormalisable supersymmetric noncommutative Wess-Zumino \cite{Girotti:2000gc} and most likely to consistent renormalisable, or even UV finite, supersymmetric noncommutative $U(N)$ theories \cite{Jack:2001jh, Ferrari:2004ex}. A noncommutative extension of the MSSM has been put forward in ref.\cite{Arai:2006ya}, which contains more ``particle'' states than the ordinary MSSM due to the noncommutative anomaly cancellation conditions \cite{GraciaBondia:2000pz, Bonora:2000he} and other noncommutative requirements.

	On the other hand, concerning the theories defined by means of Seiberg-Witten maps, they are known to have gauge anomaly cancellation conditions identical to their commutative counterparts \cite{Brandt:2003fx}, and their renormalisability properties  have been studied in a wide number of papers \cite{Bichl:2001cq,Wulkenhaar:2001sq, Buric:2002gm, Buric:2004ms, Buric:2005xe,Buric:2006wm,Martin:2006gw,Martin:2007wv,Buric:2007ix}. The results can be summarised as follows: pure gauge theories, U(1)  or SU(N), are one-loop renormalisable at least to first order in the noncommutativity parameters. The introduction of matter fields in the form of Dirac fermions or complex scalars in arbitrary representations (but such that the matter Lagrangian in terms of noncommutative fields does not involve a covariant derivative  with a star-product commutator), does not spoil the renormalisability of the gauge sector of the theory; however, the full theory seems to be nonrenormalisable in all cases analysed. These cases for which the renormalisability of the matter sector has been addressed  are: Dirac fermions with gauge groups U(1) \cite{Wulkenhaar:2001sq,Buric:2002gm} or SU(2) in the fundamental representation \cite{Buric:2004ms}, and U(1) complex scalars \cite{Martin:2006gw}. Renormalisability is spoilt by the appearance of divergences in matter field Green functions which cannot be removed by multiplicative renormalisations or field redefinitions. There is still no definitive answer concerning whether other types of matter fields or representations could overcome this problem, despite promising results concerning chiral fermions \cite{Buric:2007ix}. Still, the renormalisability properties of theories with Majorana fermions or/and covariant derivatives involving a star-product commutator have not been studied. Moreover, supersymmetry could be expected to make some divergences go away. However, though generally supersymmetry is associated with a cancellation of divergences between bosonic and fermionic degrees of freedom, and noncommutative U(N) theories defined by means of Seiberg-Witten maps have been shown to be compatible with supersymmetry, it turns out that the latter is realised nonlinearly in the ordinary fields \cite{Martin:2008xa}, and thus it is not clear how it will affect divergences.
	
	Comparing the quantum properties of noncommutative theories in both the standard and enveloping algebra approaches raises interesting questions regarding their equivalence for U(N) gauge groups, for which the Seiberg-Witten map establishes a classical equivalence. The different gauge anomaly cancellation conditions makes this equivalence doubtful in the presence of chiral fermions, at least when noncommutativity is treated perturbatively. In the case of theories without matter,  the equivalence has been found to hold for noncommutative Chern-Simons \cite{Kaminsky:2003qq} --a theory which is UV finite--, whereas for other gauge theories with or without matter there is no concluding evidence, since on the side of the enveloping algebra approach the theories studied have exclusively U(1) and SU(N) gauge groups, while to make contact with the standard formalism one should consider U(N) in the large N limit, in which the theories, at least at the one-loop level,  are supposed to be well behaved and renormalisable for infinitesimal noncommutativity.
	
	We have so far identified several issues that needed further investigation. On one hand,  the  renormalisability properties, both for the gauge sector and the full theory, of noncommutative theories defined by means of Seiberg-Witten maps with Majorana fermions and/or involving a covariant derivative with star-product commutators and/or supersymmetry. On the other hand, the equivalence at the quantum level of the standard and enveloping algebra approaches for supersymmetric noncommutative U(N) gauge theories in the large N limit,i.e., the quantum duality of supersymmetric noncommutative U(N) formulated in terms
of noncommutative fields and the supersymmetric theory, whose fields are ordinary gauge fields carrying a nonlinear realisation of supersymmetry, obtained from the former by using the Seiberg-Witten map.
	
	The aim of this paper is to address some of the open issues mentioned earlier by analysing the renormalisability properties of ${\cal N}=1$ U(N) super Yang-Mills in the enveloping algebra approach, with the ordinary fields taking values in the fundamental representation of the gauge group. First, the theory has a Majorana fermion with a covariant derivative involving  a star-product commutator; supersymmetry is also present for the noncommutative fields, and it is inherited by the ordinary fields albeit in a nonlinear fashion. Secondly, since we have a U(N) gauge group in the fundamental representation, the theory can also be formulated in the standard approach, in which case,  in the large N limit, it is renormalisable and well-behaved for small noncommutativity. We will  analyse whether one-loop renormalisability in the background field gauge is achieved at least for large N. Further, in order to complement previous research regarding theories with simple gauge groups, we will study the renormalisability properties of the SU(N) model that results from eliminating the U(1) degrees of freedom in the U(N) theory, with the goal  of seeing whether the modified field content and interactions yield a better behaviour at the quantum level. To tackle these problems, we will compute the divergent part of the one-loop effective action at first order in the noncommutative parameters $\th^{\m\n}$, using  the background field method in the background field gauge and dimensional regularisation, and we will study whether the divergences can be removed by appropriate multiplicative renormalisations of the parameters of the theory plus nonmultiplicative field redefinitions.
	
	The paper is organised as follows. The model and the background field method are introduced in section 2. Section 3 is devoted to the computation of the  full divergent part of the one-loop effective action: first, a method is outlined  which allows to obtain the full result by calculating a minimum number of diagrams, whose divergent parts are then computed in dimensional regularisation; following this the full gauge invariant expression is finally reconstructed. The renormalisability of the theory, both for arbitrary finite and large N, is studied in section 4, and then  conclusions are drawn in section 5. Two appendices are included, the first one with some Lie and Dirac algebra identities, and the second one displaying the Feynman rules employed in the computation.

	
\section{The model and the background field method}

	The action of the model, in terms of noncommutative fields, is the following,
\begin{equation}
\label{Snc}
S=\idx \!-\frac{1}{2g^2}\Tr F_{\mu\nu}\star F^{\mu\nu}+\frac{i}{g^2}\Tr\bar{\Lambda}\Dirac_\star\Lambda,\quad F_{\m\n}=\partial_\m A_\n-\partial_\n A_\m-i[A_\m, A_\n]_\star,\,D_{\star,\m}=\partial_\m-i[A_\m,\,]\star,
\end{equation}
where the fields take values in the enveloping algebra of U(N), $A_\m=A_\m^A\, T^A,$ $\Lambda=\Lambda^A T^A$ and $\Lambda$ is a Majorana spinor (see appendix A for conventions). The U(N) fields will be taken in the fundamental representation. The noncommutative product $\star$ is the usual Moyal product,
\begin{align*}
a\star b=a \exp\Big[\frac{ih}{2}\th^{\m\n}\overleftarrow{\partial}_\m\overrightarrow{\partial}_\n\Big]b,
\end{align*}
with $h$ setting the noncommutative scale. The model has ${\cal N}=1$ supersymmetry in terms of the noncommutative fields; it can be formulated in terms of a noncommutative vector superfield in the Wess-Zumino gauge.

	The noncomutative fields are defined in terms of U(N) Lie algebra valued ordinary fields, which we denote by ${\mathfrak a}_\m,{\mathfrak l},$ by means of the following Seiberg-Witten maps,
\begin{align}
\nonumber A_\m&={\mathfrak a}_\m-\frac{h}{4}\th^{\a\b}\{{\mathfrak a}_\a,\partial_\b {\mathfrak a}_\m+{\mathfrak f}_{\b\m}\}+h{\mathfrak S}_\m+O(h^2),\\
\label{SWmaps}\Lambda=&{\mathfrak l}-\frac{h}{4}\th^{\a\b}\{{\mathfrak a}_\a,2D_\b {\mathfrak l}+i[{\mathfrak a}_\b,{\mathfrak l}]\}+h{\mathfrak L}+O(h^2),
\end{align}
where $D_\m=\partial_\m-i[{\mathfrak a}_\m,\,],\,{\mathfrak f}_{\m\n}=\partial_\m{\mathfrak a}_\n-\partial_\n{\mathfrak a}_\m-i[{\mathfrak a}_\m,{\mathfrak a}_\n],$ and ${\mathfrak S}_\m,$ ${\mathfrak L}$ represent the ambiguities in the map at order $h,$ given by sums of terms  which involve a contraction with $\th^{\m\n},$ have the appropriate mass dimensions and transform in the adjoint representation of the gauge group; they can be argued to be equivalent to field redefinitions, as will be seen in section 4.

We will work with the following decomposition of the U(N) fields in the fundamental representation into their SU(N) and U(1) parts:
\begin{align}
\nonumber
 {\mathfrak a}_\m&=a_\m^a T^a+b_\m\frac{\unit}{\sqrt{2N}},\quad {\mathfrak f}_{\m\n}=f_{\m\n}^a T^a+g_{\m\n}\frac{\unit}{\sqrt{2N}},\\
 \label{decomp}{\mathfrak l}&=\la^a T^a+u\frac{\unit}{\sqrt{2N}}.
\end{align}
This will allow us to study the properties of both the U(N) theory and  the SU(N) theory that results from suppressing the U(1) degrees of freedom $b_\m,u$.

	We will argue in the next section that, for the purpose of checking renormalisability, it suffices to compute the divergent part of the effective action ignoring  at tree-level the ambiguities ${\mathfrak S}_\m,{\mathfrak L}$ of the Seiberg-Witten maps in eq.~\eqref{SWmaps}; the ambiguities, however, have to be taken into account when considering the allowed counterterms. The action in terms of ordinary fields, after expanding \eqref{Snc} with eqs.~\eqref{SWmaps} with ${\mathfrak S}_\m={\mathfrak L}=0,$ turns out to be the following
\begin{align}
\nonumber S=&S^{(0)}+hS^{(1)}+O(h^2),\\
\label{Sord}S^{(0)}=& -\frac{1}{2g^2}\idx\Tr \fk_{\mu\nu}\fk^{\mu\nu}+\frac{i}{g^2}\idx\Tr\bar{\lk}\Dirac\lk,\\
\nonumber S^{(1)}=&\frac{1}{4g^2}\idx\Tr\th^{\a\b}\fk_{\m\n}\fk^{\m\n}\fk^{\a\b}-\frac{1}{g^2}\idx\Tr\th^{\a\b}\fk_{\a\m}\fk_{\b\n}\fk^{\a\b}-\frac{i}{4}\idx\Tr\th^{\a\b}\bar\lk\ga^\m\{D_\m \lk,\fk_{\a\b}\}\\
 \nonumber &-\frac{i}{2}\idx\Tr\th^{\a\b}\bar\lk\ga^\m\{D_\b \lk,\fk_{\m\a}\}.
\end{align}
	In the previous action, all the noncommutative terms involve traces of the type $\Tr T^A\{T^B,T^C\}=\frac{1}{2}d^{ABC}$ (see appendix A). For $N<3$, the SU(N) part of the Lie algebra, for arbitrary representations, has $d^{abc}=0$, which means that the SU(N) theory obtained by eliminating the U(1) degrees of freedom is, to order $h$, equivalent to its commutative limit. Therefore, when studying the SU(N) theory we will only consider $N\geq3$. As shown in ref. \cite{Martin:2008xa} (see also \cite{Dayi:2003ju}) the fields in the action in eq.\eqref{Sord} carry a nonlinear realisation of ${\cal N}=1$ supersymmetry which define supersymmetry transformations that leave that action invariant.
	
	In the enveloping algebra approach, quantisation is performed on the ordinary fields. In order to compute the effective action with the background field method \cite{Abbott:1980hw}, we split the gauge field $\ak_\m$ in a background part $b_\m$ and a quantum part $q_\m,$
\begin{align}
\label{splitting}
\ak_\mu=b_\m+q_\m.
\end{align}
	A gauge transformation of $\ak_\m,$ $\delta \ak_\m=D_\m c,$ can be generated by two types of transformations of the fields $b,q$:
\begin{align}
\label{quantumgauge}&\text{Quantum gauge transformations: }\delta q_\m=D[q]_\m c,\,\delta b_\m=-i[b_\m,c],\,\,D[q]_\m=\partial_\m-i[q_\m,\,],\\
\label{backgroundgauge}&\text{Background gauge transformations: }\delta q_\m=-i[q_\m,c],\,\delta b_\m=D[b]_\m c,\,\,D[b]_\m=\partial_\m-i[b_\m,\,].
\end{align}
	In order to quantise $q$ with the path integral formalism, a gauge fixing procedure is needed for the transformations in eq.~\eqref{quantumgauge}. The background field method relies in a clever choice of the gauge-fixing function which is covariant under the transformations \eqref{backgroundgauge}.  With the gauge-fixing choice $G=D^{[b]}_\m q^\m=0,$ the gauge-fixing and ghost action are the following
\begin{align}\label{Sgf}
S_{gf}=-\frac{1}{2\alpha}\idx(D^{[b]}_\m q^\m)^2,\quad S_{gh}=\idx \bar cD^{[b]}_\m D^{[b+q]\m} c.
\end{align}	
	Quantising the fields $q_\m,{\lk},{\bar\lk},$ the generating functional of the background Green functions is given by
\begin{align}\label{Zbackground}
\tilde Z[\tilde J,\tilde\sigma,\tilde{\bar\sigma};b]=\int[dq][d\lk][d\bar\lk]\exp[i (S[b+q,\lk,\bar\lk]+S_{gf}[q;b]+S_{gh}[c,\bar c,q;b]+\tilde J_\m q^\m+\tilde\sigma\lk+\bar\lk\tilde{\bar\sigma})],
\end{align}
where $\tilde{J},\tilde{\s},\tilde{\bar\s}$ are  sources for the gauge field and Majorana fermions.  Note the use of ``$\tilde{}$'' to distinguish the background currents and functional generator $\tilde Z$ from the ones defining the true Green functions of the theory, when the splitting of eq.~\eqref{splitting} is not used and functional integration is performed over $\ak$. The generator of connected background Green functions is given by
\begin{align*}
	\tilde W[\tilde J,\tilde \sigma,\tilde{\bar\sigma};b]=-i{\rm ln} \tilde Z[\tilde J,\tilde \sigma,\tilde{\bar\sigma};b].
\end{align*}
Defining the background classical fields as
\begin{align*}
\tilde q=\frac{\delta \tilde W}{\delta\tilde J},\quad\tilde\lk=\frac{\delta\tilde W}{\delta\tilde \s},\quad\tilde{\bar\lk}=-\frac{\delta\tilde W}{\delta \tilde{\bar\s}},
\end{align*}
then by performing a Legendre transformation we get the functional $\tilde\Gamma$ which generates the  1PI connected background Green functions:
\begin{align}
\tilde\Gamma[\tilde q,\tilde\lk,\tilde{\bar\lk};b]=\tilde W[J,\tilde\sigma,\tilde{\bar\sigma};b]-\idx \tilde J_\m \tilde q^\m-\idx\tilde\sigma\tilde\lk-\idx\tilde{\bar\lk}\tilde{\bar\s}.
\label{tildeGamma}
\end{align}
	In a similar fashion, without using the splitting of eq.~\eqref{splitting}, one can define the true Green function generators $Z[J,\sigma,\bar\sigma]$ and $W[J,\sigma,\bar\sigma]$ as well as the true classical fields $\hat \ak,\hat \lk,\hat{\bar\lk}$.
	Standard formal manipulations show that the effective action of the theory $\Gamma[\hat \ak,\hat \lk,\hat{\bar\lk}]$ is related to $\tilde\Gamma[\tilde q,\tilde\lk,\tilde{\bar\lk};b]$ of eq.~\eqref{tildeGamma} by the following identity \cite{Abbott:1980hw}:
\begin{align}
\Gamma[\hat \ak,\hat \lk,\hat{\bar\lk}]= \tilde\Gamma[0,\tilde\lk,\tilde{\bar\lk};b]|_{b=\hat \ak,\tilde\lk=\hat\lk,\tilde{\bar\lk}=\hat{\bar\lk}},
\label{Gammas}
\end{align}
where $\Gamma$ is computed with an unusual gauge-fixing. From the r.h.s. of eq.~\eqref{Gammas} it is clear that the effective action is obtained by  calculating the background effective action for the Majorana fields after integrating out the quantum fields $q,$ with the background fields $b_\m$ taken as external sources. We thus can write
\begin{align*}
\Gamma[\hat \ak,\hat \lk,\hat{\bar\lk}]=\idx\sum_k \frac{-i}{2^k (k!)^2}\tilde\Gamma[\hat \ak]^{(k)}_{\scriptsize\begin{array}{ll}
i_1,..,i_k, & j_1,..,j_k,\\
A_1,..,A_k, & B_1,..,B_k
\end{array}}\prod_{l=1}^k\hat{\bar\lk}_{i_l}^{A_l}\prod_{p=1}^k\hat\lk_{j_p}^{B_p} ,
\end{align*}
where the factor $(k!^2)$ takes into account the permutations of the $\lk's$ and $\bar\lk's,$ while the factor $2^k$ comes from the fact that, since the Majorana fermions are self-conjugate, it is always possible to interchange one $\lk$ with an $\bar\lk$. $\tilde\Gamma[\hat\ak]^{(k)}$ is nothing but the sum of background 1PI diagrams with $k$ fermionic legs, $k$ anti-fermionic legs  and no quantum gauge field legs, and with the background field $b$ renamed as $\hat \ak$. Expanding $\tilde\Gamma[\hat \ak]^{(k)}$ in the number of background gauge fields, one gets
\begin{align}
\Gamma[\hat \ak,\hat \lk,\hat{\bar\lk}]=\idx\sum_k\sum_n \frac{-i}{2^k (k!)^2}\tilde\Gamma^{(n,k)}_{\scriptsize\begin{array}{lll}
i_1,..,i_k, & j_1,..,j_k, & \mi,..,\mu_n\\
A_1,..,A_k, & B_1,..,B_k & C_1,..,C_n
\end{array}}\prod_{l=1}^k\hat{\bar\lk}_{i_l}^{A_l}\prod_{p=1}^k\hat\lk_{j_p}^{B_p}\prod_{m=1}^n\hat \ak_{\m_m}^{C_m} .
\label{Gammaexp}
\end{align}
In the previous formula $\tilde\Gamma^{(n,k)}$ is equivalent to a background 1PI diagram with $n$ background gauge field legs,  $k$ fermionic legs and $k$ anti-fermionic legs. Note that our definitions do not involve any symmetrisation over the background gauge fields. Symmetrising over them we can make contact with the usual expansion of the effective action in terms of 1PI Green functions:
\begin{align*}
\Gamma[\hat \ak,\hat \lk,\hat{\bar\lk}]=\idx\sum_k\sum_n \frac{-i}{n!2^k (k!)^2}\Gamma^{(n,k)}_{\scriptsize\begin{array}{lll}
i_1,..,i_k, & j_1,..,j_k, & \mi,..,\mu_n\\
A_1,..,A_k, & B_1,..,B_k & C_1,..,C_n
\end{array}}\prod_{l=1}^k\hat{\bar\lk}_{i_l}^{A_l}\prod_{p=1}^k\hat\lk_{j_p}^{B_p}\prod_{m=1}^n\hat \ak_{\m_m}^{C_m} ,
\end{align*}
where $\Gamma^{(n,k)},$ which is obtained from $\tilde \Gamma^{(n,k)}$ by summing over the permutations of the background gauge fields, is the 1PI Green function with $n$ gauge fields and $k$ fermion  pairs.

	 The advantage of using background diagrams coming from the functional generator in eq.~\eqref{Zbackground} is that $\tilde\Gamma[0,\tilde\lk,\tilde{\bar\lk};b]$ is gauge invariant, so that the effective action $\Gamma[\hat \ak,\hat \lk,\hat{\bar\lk}]$ is indeed gauge invariant. As explained in the next section, this can be used to simplify the computation of the divergent part of the effective action.

	
\section{Computation of the divergent part of the effective action}

	The aim of this section is to compute the divergent part of the effective action at first order in $h\th,$ by calculating the background 1PI diagrams $\tilde\Gamma^{(n,k)}$ with no external quantum gauge fields of eq.~\eqref{Gammaexp} using the Feynman rules associated with the functional generator in eq.~\eqref{Zbackground}. These rules can be derived from the expressions for the action, gauge fixing and ghost terms given in eqs.~\eqref{Sord}, \eqref{Sgf}, keeping in mind the splitting \eqref{splitting}.
	
	Before plunging into the computation, we will justify a number of simplifications that do not imply a loss of generality on the final result concerning the regularisation and renormalisation of the theory.
\begin{itemize}

\item We shall carry out our computations in dimensional regularisation with $D=4-2\epsilon$ --it is always advisable to keep an eye on dimensional reduction. That this regularisation does not preserve supersymmetry will have no bearing on our conclusions since our computations are one-loop and the inclusion of the $\epsilon$-scalars of dimensional reduction
 to turn our dimensionally
regularised theory into a theory regularised by dimensional reduction --and thus supersymmetric-- will not modify the value of UV divergences that we will compute, but will add new ones which would be subtracted by introducing counterterms made out of  ``evanescent'' operators and couplings --see
ref.\cite{Capper:1979ns, Jack:1997sr} for further details.

\item Choice of gauge $\alpha=1$ in the gauge-fixing term in eq.\eqref{Sgf}. This choice of gauge simplifies the gauge field propagator.  This brings up the question of whether, if problematic divergences appear for $\alpha=1$ that make the theory nonrenormalisable, the consideration of an arbitrary $\alpha$ might help remove these divergences. The answer is negative whenever any of the problematic divergences appearing at $\alpha=1$  do not go away on the mass shell. This is due to the results in ref.\cite{Fradkin:1983nw}
(see also \cite{Rebhan:1986wp}) which establish that the background field effective action is independent of the gauge-fixing term if the background fields are on shell.
Thus, when the background fields are on shell any  divergent contribution remaining will be independent of any gauge-fixing term 
that we chose.

\item Setting  to zero the tree-level ambiguities ${\mathfrak S}_\m,{\mathfrak L}$ of the Seiberg-Witten map of eq.~\eqref{SWmaps}. This choice  simplifies greatly the computation of the diagrams, though when studying renormalisability one can still contemplate infinite renormalisations of  ${\mathfrak S}_\m,{\mathfrak L},$  which tantamounts to consider the most general field redefinitions that cannot be reabsorbed by gauge transformations, as will be explained in section 4. Again, one may still object that considering arbitrary ${\mathfrak S}_\m,{\mathfrak L}$ at tree level might be of use to cancel possible pathological divergences (i.e., that cannot be removed by field redefinitions or multiplicative renormalisation) appearing for ${\mathfrak S}^{\rm tree}_\m={\mathfrak L}^{\rm tree}=0$. This possibility is precluded  by the arguments presented in ref.~\cite{Bonneau:1985ea}, proven there for a specific model but expected to have general validity. In this reference the authors claim that, given a theory which is multiplicatively renormalisable, then by quantising the theory after performing a field redefinition, the divergences in terms of the new fields can be reabsorbed by  the same multiplicative renormalisations of physical parameters as in the original case, plus infinite field redefinitions. In our case, we worry about possible divergences at order $h\theta$ for ${\mathfrak S}^{\rm tree}_\m={\mathfrak L}^{\rm tree}=0$ which cannot be removed by infinite field redefinitions. The theory at order $h^0$ is known to be multiplicatively renormalisable, and considering arbitrary  ${\mathfrak S}^{\rm tree}_\m,{\mathfrak L}^{\rm tree}$ is equivalent to performing finite field redefinitions of order $h$ on the ordinary fields $\ak_\m,\lk,\bar\lk$. Thus, the additional divergences dependent on  ${\mathfrak S}^{\rm tree}_\m,{\mathfrak L}^{\rm tree}$ that might appear would be equivalent to infinite field redefinitions and therefore by assumption would not be useful to cancel the original problematic divergences at ${\mathfrak S}^{\rm tree}_\m={\mathfrak L}^{\rm tree}=0.$  It follows that the conclusions about the renormalisability of the theory obtained for ${\mathfrak S}^{\rm tree}_\m={\mathfrak L}^{\rm tree}=0$ have a general validity.

\item Computing a minimum number of diagrams. The use of the background field method guarantees that the result for the effective action will be gauge invariant. Furthermore, its divergent part computed in dimensional regularisation will be local. Thus, if one chooses a basis of all  possible local gauge invariant terms up to order $h,$ the divergent part of the effective action will be a linear combination of these terms. The coefficients in this linear combination can be determined by identifying its contributions with any given number and types of fields with the poles in the dimensional regularisation parameter $\epsilon$ of the corresponding 1PI Green functions with the same number and types of external fields. By appropriately choosing the basis, it can be guaranteed that the contributions to its elements with a minimum number of fields are also independent of each other, so that the unknown coefficients in the expansion of the divergent part of the effective action in terms of the basis can be determined from the diagrams with lowest number of fields.
\end{itemize}
	
We have thus argued that we can determine unambiguously the renormalisability of the theory by computing the effective action for $\alpha=1,$ ${\mathfrak S}^{\rm tree}_\m={\mathfrak L}^{\rm tree}=0$. Under these assumptions, the Feynman rules relevant to our computations  are those given in appendix B; they use a compact notation for the Lie algebra indices, following ref.~\cite{deAzcarraga:1997ya, Bonora:2000ga}, in which the U(N) field expansion in the Lie algebra generators in the fundamental representation is taken as
\begin{align*}
{\mathfrak a}_\mu={\mathfrak a_\m}^A T^A,
\end{align*}
where $T^A=\{T^0,T^a\},$ with $T^0=\frac{\unit}{\sqrt{2N}}$  the U(1) generator and  $T^a$ denoting the SU(N) generators; more details are given in appendix A. This allows to compute simultaneously diagrams involving both SU(N) and U(1) fields, and the results for the SU(N) theory can also be easily obtained by setting the external ``$A$" indices to SU(N) indices ``$a$", and by taking care to drop the contributions of U(1) indices in terms involving contractions of internal U(N) Lie algebra indices ``$A$".

Let us start by identifying the diagrams that need to be computed by constructing the appropriate basis of local gauge invariant terms whose integrals are independent. We use the decomposition in eq.~\eqref{decomp}. Local gauge invariant terms are then constructed from traced products of the field strengths and fermion fields and their covariant derivatives; we can classify them in three sectors: SU(N) sector -only including fields in the Lie algebra of SU(N)- U(1) sector, and mixed sector. A list follows:

\noindent SU(N) sector:
\begin{align}
\nonumber t_1&=\th^{\a\b}\Tr f_{\a\b}f_{\m\n}f^{\m\n}, & t_2&=\th^{\a\b}\Tr f_{\a\m}f_{\b\n}f^{\m\n},\\
 \nonumber t_3&=\th^{\a\b}\Tr\bar\la\ga_\a D^2D_\b\la, &t_4&=\th^{\a\b}\Tr\bar\la{\ga_{\a\b}}^\m D^2D_\m\la, \\
 \nonumber t_5&=\th^{\a\b}\Tr\bar\la\ga^\m\{f_{\m\b},D_\a\la\}, &t_6&=\th^{\a\b}\Tr\bar\la\ga^\m\{f_{\a\b},D_\m\la\},\\
\label{ts} t_7&=\th^{\a\b}\Tr\bar\la\ga_\a\{f_{\b\m},D^\m\la\},& t_8&=\th^{\a\b}\Tr\bar\la{\ga_{\a\b}}^\m\{D^\n f_{\m\n},\la\}, \\
 \nonumber t_9&=\th^{\a\b}\Tr\bar\la{\ga_{\a}}^{\r\s}\{D_\b f_{\r\s},\la\}, &t_{10}&=\th^{\a\b}\Tr\bar\la{\ga}^\m[D_\m f_{\a\b},\la],\\
 \nonumber t_{11}&=\th^{\a\b}\Tr\bar\la{\ga}_\a[D^\m f_{\b\m},\la],& t_{12}&=\th^{\a\b}\Tr\bar\la{\ga_{\a\b}}^\m[f_{\m\n},D^\n\la],\\
  \nonumber t_{13}&=\th^{\a\b}\Tr\bar\la{\ga_{\a}}^{\r\s}[f_{\r\s},D_\b\la],& t_{14}&=\th^{\a\b}\Tr\bar\la{\ga_{\a}}^{\r\s}[f_{\b\s},D_\r\la],\\
 \nonumber t_{15}&=\th^{\a\b}\Tr\bar\la_i(\ga_\a)_{ij}[\{\bar\la_k,\!(\ga_\b\la)_k\},\la_j], & t_{16}&=\th^{\a\b}\Tr\bar\la_i(\!\ga^\m)_{ij}[[\bar\la_k,\!(\!\ga_{\m\a\b}\la)_k],\!\la_j].
 \end{align}
 U(1) sector:
 \begin{align}
\nonumber  u_1&=\th^{\a\b}g_{\a\b}g^{\m\n}g_{\m\n}, & u_2&=\th^{\a\b}g_{\a\m}g_{\b\n}g^{\m\n}, & u_3&=\th^{\a\b}\bar u \ga_\a\partial^2\Db u,\\
\label{us} u_4&=\th^{\a\b}\bar u{\ga_{\a\b}}^\m \partial^2\Dm u, & u_5&=\th^{\a\b}\bar u \ga^\m \Da u g_{\m\b}, & u_6&=\th^{\a\b}\bar u \ga^\m \Dm u g_{\a\b},\\
 \nonumber u_7&=\th^{\a\b}\bar u \ga_\a \partial^\m u g_{\b\m}, & u_8&=\th^{\a\b}\bar u {\ga_{\a\b}}^\m u\partial^\n g_{\m\n}, & \nonumber u_9&=\th^{\a\b}\bar u{\ga_\a}^{\r\s}u\Db g_{\r\s}.
  \end{align}
Mixed sector:
\begin{align}
\nonumber v_1&=\th^{\a\b}\Tr g_{\a\b}f^{\m\n}f_{\m\n}, & v_2&=\th^{\a\b}\Tr g^{\m\n}f_{\a\m}f_{\b\n}, &v_3&= \th^{\a\b}\Tr g_{\a\m}f_{\b\n}f^{\m\n},\\
\nonumber  v_4&=\th^{\a\b}\Tr g_{\m\n}f_{\a\b}f^{\m\n}, & v_5&=\th^{\	a\b}\Tr \bar u \ga^\m f_{\m\b}D_\a\la, & v_6&=\th^{\a\b}\Tr \bar u \ga^\m f_{\a\b}D_\m\la,\\
\nonumber  v_7&=\th^{\a\b}\Tr \bar u \ga^\m D_\m f_{\a\b}\la, & v_8&=\th^{\a\b}\Tr \bar u \ga_\a f_{\b\m}D^\m\la, & v_9&=\th^{\a\b}\Tr\bar u \ga_\a D^\m f_{\b\m}\la,\\
\label{vs} v_{10}&=\th^{\a\b}\Tr \bar u {\ga_{\a\b}}^\m D^\n f_{\m\n}\la, &  v_{11}&=\th^{\a\b} \Tr\bar u {\ga_{\a\b}}^\m  f_{\m\n}D^\n\la, &
 v_{12}&=\th^{\a\b} \Tr\bar u {\ga_\a}^{\r\s} D_\b f_{\r\s}\la, \\
 \nonumber  v_{13}&=\th^{\a\b} \Tr\bar u {\ga_\a}^{\r\s}  f_{\r\s}D_\b\la, &   v_{14}&=\th^{\a\b}\Tr \bar u {\ga_\a}^{\r\s}  f_{\b\s}D_\r\la, &  v_{15}&=\th^{\a\b}\Tr \bar \la \ga^\m D_\a \la g_{\m\b},\\
 \nonumber  v_{16}&=\th^{\a\b}\Tr \bar \la \ga^\m D_\m \la g_{\a\b},  & v_{17}&=\th^{\a\b}\Tr\bar \la \ga_\a D^\m \la g_{\b\m}, & v_{18}&=\th^{\a\b}\Tr \bar \la {\ga_{\a\b}}^\m  \la \partial^\n g_{\m\n},\\
\nonumber    v_{19}&=\th^{\a\b}\Tr \bar \la {\ga_\a}^{\r\s} \la \partial_\b g_{\r\s}.
\end{align}

In the formulae above, ``$\Tr$'' denotes the trace over the SU(N) generators. The list of terms spans modulo total derivatives all the possible gauge invariant terms of order $h\th^{\m\n}$ with the appropriate dimensions with zero or two Majorana fields. Again, the Majorana properties \eqref{Maj1} and \eqref{Maj2} have been used, so that any term with two Majorana fermions not present above can be expressed as a linear combination of the $t_i,u_i$ and $v_i,$ again modulo total derivatives. In the case of terms with four Majorana fermions,  $t_{15}$ and $t_{16}$ do not span all the allowed contributions, but the missing ones will play no role in our calculations and we will safely ignore them.

The contributions to the previous list of terms with a minimum number of fields are independent of each other, which, as explained before, allows to fix the coefficients of the expansion of the divergent part of the effective action, $\Gamma^{div},$ in terms of the $t_i,u_i,v_i$ by computing only the 1PI diagrams with the least possible number of fields. Let us identify the diagrams that need to be computed, using the notation in eq.~\eqref{Gammaexp} for the 1PI background Green functions.

At order $h^0,$ the possible gauge invariant terms are $\Tr f_{\m\n}f^{\m\n}$ and $\Tr\bar\la \Dirac \la$. Thus, using the notation of eq.~\eqref{Gammaexp} only the diagrams contributing to $\tilde\Gamma^{(2,0)}$ -with two external background gauge field legs- and $\tilde\Gamma^{(0,1)}$ -with two external quantum fermionic legs- need to be computed. At order $h,$ we have, schematically, the following types of terms:
\begin{itemize}
\item Terms of the type $\Tr \th fff, \Tr\th g ff, \th ggg,$ which are spanned by $t_1,t_2,$ $u_1,u_2$ and $v_1-v_4$ in eqs.~\eqref{ts} ,\eqref{us} and \eqref{vs}, whose contributions with  three gauge fields are independent.  Thus it suffices to compute diagrams with three external gauge fields, contributing to $\tilde\Gamma^{(3,0)}$.

\item Terms of the type $\Tr\th\bar\la D^3\la,\th\bar u  \partial^3u,$ which are spanned by $t_3,t_4$ and $u_3,u_4$ in eqs.~\eqref{ts} and \eqref{us}. They involve at least two fermionic fields, so that their coefficients in the expansion of $\Gamma^{div}$ can be fixed by computing $\tilde\Gamma^{(0,1)},$ which arises from diagrams with two fermionic legs.

\item Terms of the type  -neglecting ordering- $\Tr\th\bar\la Df\la,\Tr\th\bar\la f D\la,\th\bar u\partial g u, \th \bar ug \partial u,\Tr\bar u D f\la,$ $ \Tr\bar u f D\la,$ $\Tr\bar\la D\la g, \Tr\bar\la\la\partial g,$ which are spanned by $t_5-t_{14}, u_5-u_{9}, v_{5}-v_{19}$ in eqs.~\eqref{ts}, \eqref{us} and \eqref{vs}. Their contributions  with one gauge field and two Majorana fields are again independent, so that it suffices to compute the diagrams contributing to $\tilde\Gamma^{(1,1)},$ i.e., with one background gauge field leg and two quantum fermionic legs.

\item Terms of the type $\Tr\th\bar\la\la\bar\la\la,$ such as $t_{15},t_{16}$ in eq.~\eqref{ts}. Though  $t_{15},t_{16}$ do not span all possibilities, it is clear that the computation of $\tilde\Gamma^{(0,2)}$ (diagrams with four external fermionic legs) will completely determine the corresponding contribution to the effective action $\Gamma$.
\end{itemize}
Summarising, at order $h$ the only diagrams that have to be computed are those contributing to the 1PI Green functions   $\tilde\Gamma^{(3,0)},$ $\tilde\Gamma^{(0,1)},$  $\tilde\Gamma^{(1,1)}$ and  $\tilde\Gamma^{(0,2)}$. We proceed in the next sections, using dimensional regularisation at $D=4-2\epsilon$ dimensions, with the Feynman rules displayed in appendix B. The calculations are quite involved and were done with the symbolic manipulation software $\tt Mathematica$.


\subsection{Commutative limit}
Here we quote the known commutative result for the dimensionally regularised divergent part of the effective action:
\begin{align}
\label{Gammaord}
	\Gamma^{\rm ord,div}_{[U(N)]}=\iDx-\frac{3g^2N}{16\pi^2\epsilon}\Tr\Big[-\frac{1}{2g^2} f_{\m\n}f^{\m\n}]+\iDx\frac{N}{16\pi^2\epsilon}[i\Tr\bar\la\Dirac\la].
\end{align}
For simplicity, we suppressed the ``$\hat{}$" symbols with which we denoted the classical fields in section 2; we will keep doing so in the rest of the paper. Note that the divergent part only involves the SU(N) fields $a,\la,$ since the U(1) sector is free in the commutative limit. In fact, since the U(1) sector is free, in the SU(N) case the result is identical,
\begin{align}
\label{GammaordSUN}
\Gamma^{\rm ord,div}_{[SU(N)]}=\Gamma^{\rm ord,div}_{[U(N)]}.
\end{align}



\subsection{Noncommutative contributions to $\tilde\Gamma^{(3,0)}$}
\begin{figure}[h]\centering
\includegraphics[scale=1.07]{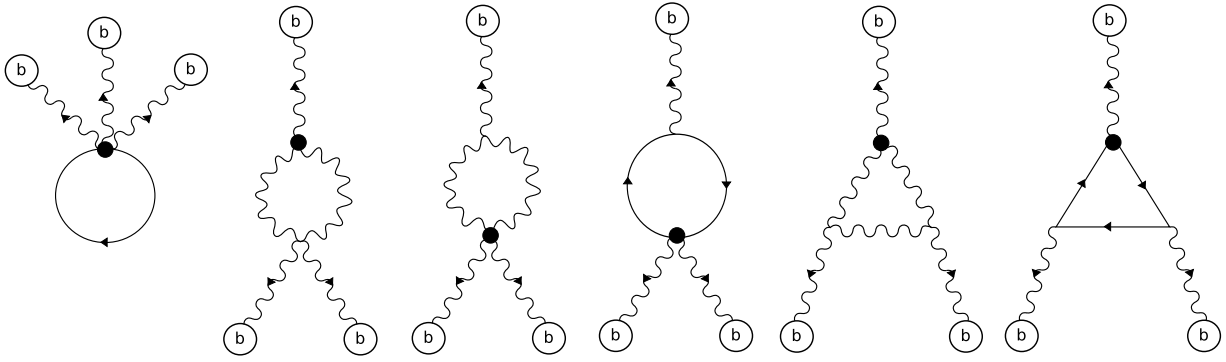}
\caption{Diagrams contributing to  $\tilde\Gamma^{(3,0)}$ at order $h$.}
\label{f:1}
\end{figure}

The diagrams that contribute are shown in Fig. \ref{f:1}.
	Note that, though we did not provide in  appendix B the Feynman rule for the vertex appearing in the first diagram, this diagram is directly zero since it involves an integral of the type
\begin{align}\label{zeroint}
\int d^Dl\frac{\prod_i l_{\mu_i}}{(l^2)^k},
\end{align}	
which vanishes in dimensional regularisation.
	
	The results for the diagrams are too lengthy to be displayed here individually. We will quote the final expression for the contribution to the divergent part of the effective action in position space:
\begin{align}
\label{Gaaa}i\tilde\Gamma^{(3,0),\rm NC,div}_{[U(N)]\scriptsize\begin{array}{l}
\mi,\mii,\miii\\
A_1,A_2,A_3
\end{array}}\ak_{\mi}^{A_1}\ak_{\mii}^{A_2}\ak_{\miii}^{A_3}=&\frac{3g^2Nh}{16\pi^2\epsilon}\Big[\frac{1}{4g^2}t_1-\frac{1}{g^2}t_2\Big]\Big|_{aaa}\\
\nonumber&+\frac{2g^2Nh}{16\pi^2\epsilon}\Big[\frac{1}{4g^2\sqrt{2N}}(v_1+2v_4)-\frac{1}{g^2\sqrt{2N}}(v_2+2v_3)\Big]\Big|_{baa}+O(h^2),
\end{align}
where ``$|_{aaa}$" and $|_{baa}$" denote the contributions with lowest number of fields, i.e., three SU(N) gauge fields and   one U(1) and two SU(N) gauge fields, respectively. Recall that the $t_i,u_i,v_i$ are the gauge invariant terms defined in eqs.~\eqref{ts}, \eqref{us} and \eqref{vs}.
To get the SU(N) result, the external Lie algebra indices of the diagrams have to be set to SU(N) indices, and any U(1) contributions to internal contractions have to be eliminated. It turns out that all diagrams involve contractions of the type appearing in eq.~\eqref{UNcont} of appendix A, which, when setting the uncontracted indices to SU(N) indices, do not involve any contributions from internal U(1) indices.  This is equivalent to saying that the U(1) fields do not run in the loops when the external fields are the $a_\m^a$. From this we conclude that the SU(N) result is obtained from eq.~\eqref{Gaaa} by simply setting to zero the U(1) fields:
\begin{align}\label{GaaaSUN}
i\tilde\Gamma^{(3,0),\rm NC,div}_{[SU(N)]\scriptsize\begin{array}{l}
\mi,\mii,\miii\\
a_1,a_2,a_3
\end{array}} a_{\mi}^{a_1} a_{\mii}^{a_2} a_{\miii}^{a_3}=&\frac{3g^2Nh}{16\pi^2\epsilon}\Big[\frac{1}{4g^2}t_1-\frac{1}{g^2}t_2\Big]\Big|_{aaa}+O(h^2).
\end{align}


\subsection{Noncommutative contributions to  $\tilde\Gamma^{(0,1)}$}

\begin{figure}[h]\centering
\includegraphics[scale=1.2]{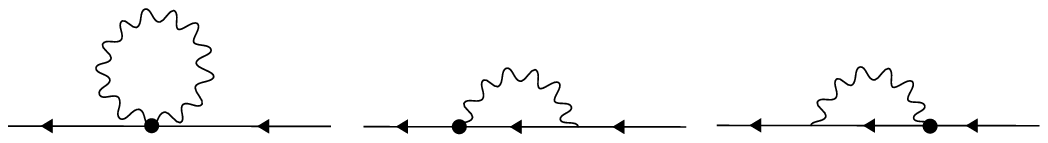}
\caption{Diagrams contributing to  $\tilde\Gamma^{(0,1)}$ at order $h$.}
\label{f:2}
\end{figure}

The diagrams contributing to the $\tilde\Gamma^{(0,1)}$ Green function at order $\theta$ are shown in Fig. \ref{f:2}.
The first diagram is zero as it involves again an integral of the type shown in eq.~\eqref{zeroint}.
For external colour indices A, B, it is easily seen that the rest of the diagrams are zero since they are proportional to either $f^{ACD}d^{BCD}=0$ or $f^{BCD}d^{ACD}=0$. To get the SU(N) result one has to set the external indices to $a,$ $b$ and drop any $U(1)$ contributions in the contractions of the internal indices. However, since $f^{bCD}d^{aCD}=f^{bcd}d^{acd},$ no U(1) contributions must be eliminated, and the same argument as before applies. Therefore,
\begin{align}\label{Gaa}
\tilde\Gamma^{(0,1),\rm NC,div}_{[U(N)]}=\tilde\Gamma^{(0,1),\rm NC,div}_{[SU(N)]}=O(h^2).
\end{align}


\subsection{Noncommutative contributions to  $\tilde\Gamma^{(1,1)}$ }
\begin{figure}[h]\centering
\psfrag{B1}{$B_1$}\psfrag{B2}{$B_2$}\psfrag{B3}{$B_3$}\psfrag{B4}{$B_4$}\psfrag{B5}{$B_5$}\psfrag{B6}{$B_6$}
\psfrag{B7}{$B_7$}\psfrag{B8}{$B_8$}\psfrag{B9}{$B_9$}\psfrag{B10}{$B_{10}$}
\psfrag{p}{$p$}\psfrag{q}{$q$}\psfrag{k}{$k$}
\psfrag{j}{$j,B$}\psfrag{i}{$i,A$}\psfrag{m}{$\mu,C$}
\includegraphics[scale=1.2]{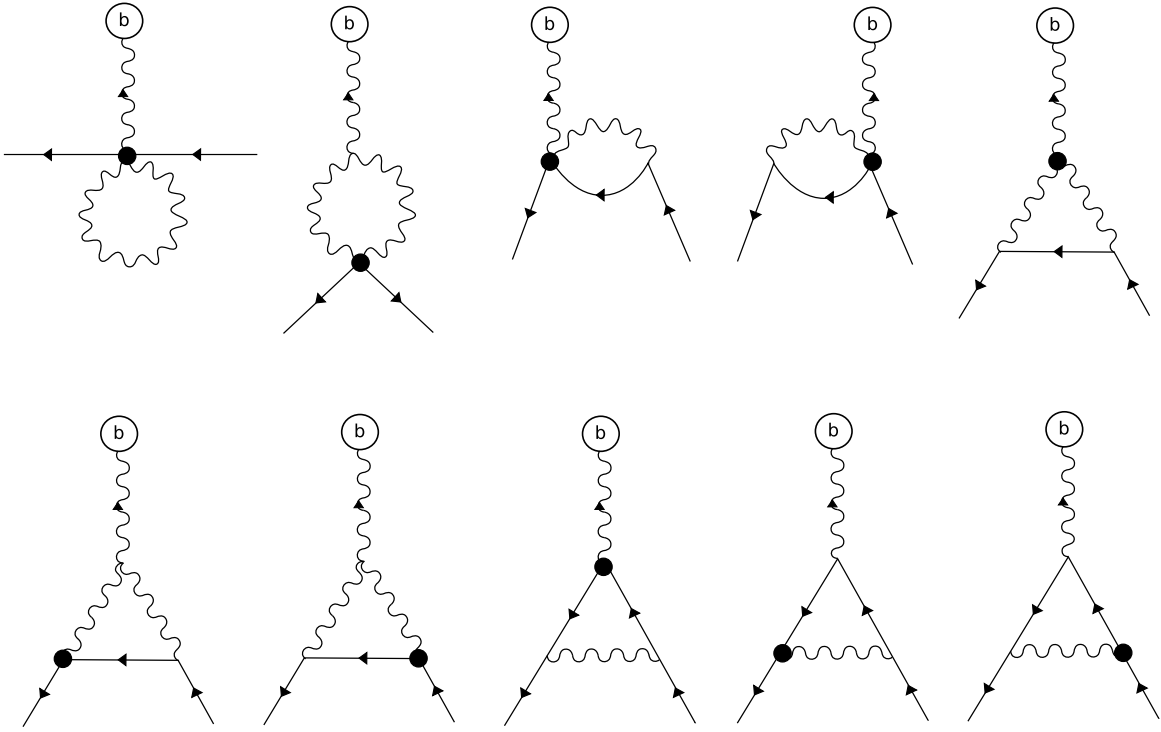}
\caption{Diagrams contributing to  $\tilde\Gamma^{(1,1)}$ at order $h$.}
\label{f:3}
\end{figure}

The diagrams contributing to the $\tilde\Gamma^{(1,1)}$ Green function at order $\theta$ are shown in Fig. \ref{f:3}.
Again, we will write down the final result of the lengthy computation:
\begin{align}
\nonumber\frac{i}{2}\tilde\Gamma^{(1,1),\rm NC,div}_{[U(N)]\scriptsize\begin{array}{l}
i, j, \m\\
A, B, C
\end{array}}{\bar\lk}^A_i \lk^B_j \ak_{\m}^{C}=&-\frac{iNh}{16\pi^2\epsilon}\Big[\frac{1}{4}t_6-\frac{1}{2}t_7-\frac{1}{8}t_8-\frac{1}{16}t_9\Big]\Big|_{a\bar\la\la}\\
\label{Gall}&+\frac{iNh}{16\pi^2\epsilon}\Big(\frac{1}{\sqrt{2N}}\Big)\Big[v_5-\frac{3}{2}v_6+2v_8-\frac{1}{4}v_{10}-\frac{1}{2}v_{12}\Big]\Big|_{a\bar u\la}\\
\nonumber &+\frac{iNh}{16\pi^2\epsilon}\Big(\frac{1}{\sqrt{2N}}\Big)\Big[-v_{15}+\frac{1}{2}v_{16}+\frac{3}{4}v_{18}+\frac{3}{4}v_{19}\Big]\Big|_{b\bar\la\la}+O(h^2).
\end{align}
To get the SU(N) result, using the same arguments as in the previous subsection it suffices to set the U(1) fields to zero:
\begin{align}
\frac{i}{2}\tilde\Gamma^{(1,1),\rm NC,div}_{[SU(N)]\scriptsize\begin{array}{l}
i, j, \m\\
a, b, c
\end{array}}\bar\la^a_i \la^b_j a_{\m}^{c}=&-\frac{iNh}{16\pi^2\epsilon}\Big[\frac{1}{4}t_6-\frac{1}{2}t_7-\frac{1}{8}t_8-\frac{1}{16}t_9\Big]\Big|_{a\bar\la\la}+O(h^2).\label{GallSUN}
\end{align}


\subsection{Noncommutative contributions to $\tilde\Gamma^{(0,2)}$}
\begin{figure}[h]\centering
\psfrag{C1}{$C_1$}\psfrag{C2}{$C_k,k=1...16$}\psfrag{p}{$p$}\psfrag{q}{$q$}\psfrag{r}{$r$}\psfrag{s}{$s$}
\psfrag{j}{$j,B$}\psfrag{i}{$i,A$}\psfrag{k}{$k,C$}\psfrag{l}{$l,D$}
\psfrag{per}{{$+$ three perm. of momenta and indices}}\psfrag{per2}{$+3$ perm.}\hskip-3cm
\includegraphics[scale=1.2]{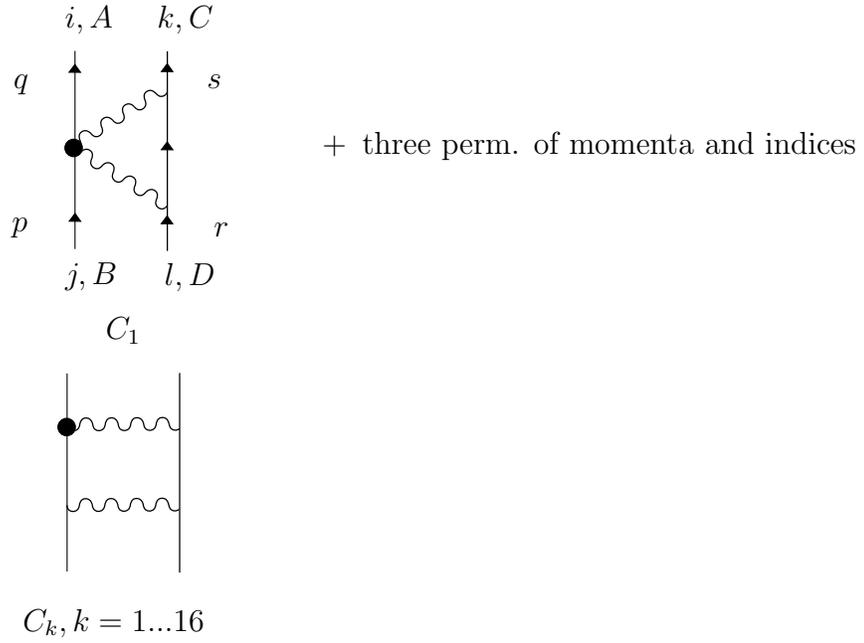}
\caption{Diagrams contributing to  $\tilde\Gamma^{(0,2)}$ at order $h$.}
\label{f:4}
\end{figure}
The diagrams that contribute are shown in Fig.~\ref{f:4}; it is easily seen that the box diagrams are finite since, though they would appear to be logarithmically divergent, one of the momenta in the noncommutative vertex is always external, as can be seen from the Feynman rule in appendix B. The final result is as follows:
\begin{align}
\label{Gllll}
&\frac{i}{16}\tilde\Gamma^{(0,2),\rm NC,div}_{[U(N)]\scriptsize\begin{array}{l}
i, j,k,l \\
A, B, C,D
\end{array}}{\bar\lk}^A_i {\bar\lk}^C_k\lk^B_j\lk^D_l=-\frac{3iNh}{512\pi^2\epsilon}t_{16}+O(h^2).
\end{align}
Again, for external SU(N) fields no U(1) fields run in the loops, and the SU(N) result is identical.
\begin{align}\label{GllllSUN}
&\frac{i}{16}\tilde\Gamma^{(0,2),\rm NC,div}_{[SU(N)]\scriptsize\begin{array}{l}
i, j,k,l \\
a, b,c,d
\end{array}}{\bar\la}^a_i {\bar\la}^c_k\la^b_j\la^d_l=-\frac{3iNh}{512\pi^2\epsilon}t_{16}+O(h^2).
\end{align}


\subsection{Final expression}

	From the previous discussions and the notation employed in the results of the 1PI Green functions in eqs.~\eqref{Gaaa}, \eqref{Gaa},  \eqref{Gall} and \eqref{Gllll}, which are expressed as the contributions with the lowest number of fields of linear combinations of the gauge invariant terms $t_i,u_i,v_i$ of eqs.~\eqref{ts}, \eqref{us} and \eqref{vs}, it is clear that the final result for the first-order noncommutative correction to the divergent part of the one-loop effective action is simply given by the integral of a sum of the $t_i,u_i,v_i$ with the same coefficients as in eqs.~\eqref{Gaa}, \eqref{Gaaa}, \eqref{Gall} and \eqref{Gllll}:
\begin{align}
\nonumber\Gamma^{\rm div,NC}_{[U(N)]}=&-\,h\iDx\Big(\frac{3g^2N}{16\pi^2\epsilon}\Big[\frac{1}{4g^2}t_1-\frac{1}{g^2}t_2\Big]+\frac{2g^2N}{16\pi^2\epsilon}\Big[\frac{1}{4g^2\sqrt{2N}}(v_1+2v_4)-\frac{1}{g^2\sqrt{2N}}(v_2+2v_3)\Big]\\
\nonumber&-\frac{iN}{16\pi^2\epsilon}\Big[\frac{1}{4}t_6-\frac{1}{2}t_7-\frac{1}{8}t_8-\frac{1}{16}t_9\Big]+\frac{iN}{16\pi^2\epsilon}\Big(\frac{1}{\sqrt{2N}}\Big)\Big[v_5-\frac{3}{2}v_6+2v_8-\frac{1}{4}v_{10}-\frac{1}{2}v_{12}\\
\label{Gammadivts}
 &-v_{15}+\frac{1}{2}v_{16}+\frac{3}{4}v_{18}+\frac{3}{4}v_{19}\Big]-\frac{3iN}{512\pi^2\epsilon}t_{16}\Big)+O(h^2).
\end{align}
Similarly, the SU(N) result obtained from the expressions in eqs.~ \eqref{GaaaSUN}, \eqref{Gaa}, \eqref{GallSUN} and \eqref{GllllSUN} is
\begin{align}
\nonumber
\Gamma^{\rm div,NC}_{[SU(N)]}=&-\,h\iDx\Big(\frac{3g^2N}{16\pi^2\epsilon}\Big[\frac{1}{4g^2}t_1-\frac{1}{g^2}t_2\Big]-\frac{iN}{16\pi^2\epsilon}\Big[\frac{1}{4}t_6-\frac{1}{2}t_7-\frac{1}{8}t_8-\frac{1}{16}t_9\Big]-\frac{3iN}{512\pi^2\epsilon}t_{16}\Big)\\
&+O(h^2).\label{GammadivtsSUN}
\end{align}

Equivalently, substituting the expressions in eqs.~\eqref{ts}, \eqref{us} and \eqref{vs}, and adding the commutative contribution of eq.~\eqref{Gammaord}, we arrive to the following formula for the one-loop divergent part of the effective action at first order in the noncommutative parameters
\begin{align}
\nonumber&\Gamma^{\rm div}_{[U(N)]}=-\iDx\Big(\frac{3g^2N}{16\pi^2\epsilon}\Tr\Big[-\frac{1}{2g^2} f_{\m\n}f^{\m\n}+\frac{h}{4g^2}\th^{\m\n} f_{\a\b}f_{\m\n}f^{\m\n}-\frac{h}{g^2}\th^{\a\b} f_{\a\m}f_{\b\m}f^{\m\n}\Big]\\
\nonumber &-\frac{N}{16\pi^2\epsilon}[i\Tr\bar\la\Dirac\la]\\
\nonumber&+\frac{2g^2Nh}{16\pi^2\epsilon}\Tr\Big[\frac{1}{4g^2\sqrt{2N}}\th^{\a\b}(g_{\a\b} f^{\m\n}f_{\m\n}+2g_{\m\n} f_{\a\b}f^{\m\n})-\frac{1}{g^2\sqrt{2N}}\th^{\a\b}(g^{\m\n} f_{\a\m}f_{\b\n}+2g_{\a\m} f_{\b\m}f^{\m\n})\Big]\\
\nonumber&-\frac{Nh}{16\pi^2\epsilon}\Tr\Big[\frac{i}{4}\th^{\a\b}\bar\la\ga^\m\{f_{\a\b},D_\m\la\}-\frac{i}{2}\th^{\a\b}\bar\la\ga_\a\{f_{\b\m},D^\m\la\}-\frac{i}{8}\th^{\a\b}\bar\la{\ga_{\a\b}}^\m\{D^\n f_{\m\n},\la\}\\
\nonumber&-\frac{i}{16}\th^{\a\b}\bar\la{\ga_\a}^{\r\s}\{D_\b f_{\r\s},\la\}\Big]+\frac{Nh}{16\pi^2\epsilon}\Big(\frac{1}{\sqrt{2N}}\Big)\Tr\Big[i\th^{\a\b}\bar u\ga^\m f_{\m\b}D_\a\la-\frac{3}{2}i\th^{\a\b}\bar u\ga^\m f_{\a\b}D_\m\la\\
\label{Gammatotaldiv}&+2i\th^{\a\b}\bar u\ga_\a f_{\b\m}D^\m\la-\frac{i}{4}\th^{\a\b}\bar u{\ga_{\a\b}}^\m D^\n f_{\m\n} \la-\frac{i}{2}\th^{\a\b}\bar u{\ga_\a}^{\r\s}D_\b f_{\r\s}\la\Big]\\
\nonumber&+\frac{Nh}{16\pi^2\epsilon}\Big(\frac{1}{\sqrt{2N}}\Big)\Tr\Big[-i\th^{\a\b}\bar \la\ga^\m D_\a\la g_{\m\b}+\frac{i}{2}\th^{\a\b}\bar \la\ga^\m D_\m\la g_{\a\b}+\frac{3i}{4}\th^{\a\b}\bar\la{\ga_{\a\b}}^\m  \la\partial^\n g_{\m\n}\\
\nonumber &+\frac{3i}{4}\th^{\a\b}\bar \la{\ga_\a}^{\r\s}\la\partial_\b g_{\r\s}\Big]-\frac{3iNh}{512\pi^2\epsilon}
\th^{\a\b}\Tr\bar\la_i(\!\ga^\m)_{ij}[[\bar\la_k,\!(\!\ga_{\m\a\b}\la)_k],\!\la_j]\Big)+O(h^2).
\end{align}
The corresponding expression in the SU(N) case is
\begin{align}
\nonumber&\Gamma^{\rm div}_{[SU(N)]}=-\iDx\Big(\frac{3g^2N}{16\pi^2\epsilon}\Tr\Big[-\frac{1}{2g^2} f_{\m\n}f^{\m\n}+\frac{h}{4g^2}\th^{\m\n} f_{\a\b}f_{\m\n}f^{\m\n}-\frac{h}{g^2}\th^{\a\b} f_{\a\m}f_{\b\m}f^{\m\n}\Big]\\
\nonumber&-\frac{N}{16\pi^2\epsilon}[i\Tr\bar\la\Dirac\la]\\
\nonumber&-\frac{Nh}{16\pi^2\epsilon}\Tr\Big[\frac{i}{4}\th^{\a\b}\bar\la\ga^\m\{f_{\a\b},D_\m\la\}-\frac{i}{2}\th^{\a\b}\bar\la\ga_\a\{f_{\b\m},D^\m\la\}-\frac{i}{8}\th^{\a\b}\bar\la{\ga_{\a\b}}^\m\{D^\n f_{\m\n},\la\}\\
\label{GammatotaldivSUN}&-\frac{i}{16}\th^{\a\b}\bar\la{\ga_\a}^{\r\s}\{D_\b f_{\r\s},\la\}\Big]
 -\frac{3iNh}{512\pi^2\epsilon}
\th^{\a\b}\Tr\bar\la_i(\!\ga^\m)_{ij}[[\bar\la_k,\!(\!\ga_{\m\a\b}\la)_k],\!\la_j]\Big)+O(h^2).
\end{align}
It is worth noting that, for $N=2$, all the terms with SU(N) fields whose traces yield factors $d^{abc}=2\Tr T^a\{T^b,T^c\}$ in eqs.~\eqref{Gammatotaldiv} and eq.~\eqref{GammatotaldivSUN} vanish. This means that all terms involving only SU(N) fields vanish; in the U(2) case, we are only left with SU(2)-U(1) mixed terms, while for the SU(2) theory the noncommutative divergences disappear. This fact is independent of the representation considered since, for a representation $R$ of $SU(N)$, $\Tr_R T^a\{T^b,T^c\}\propto Tr_F T^a\{T^b,T^c\}$.

In the U(1) case, it is also clear from eqs.~\eqref{Gammatotaldiv} and eq.~\eqref{GammatotaldivSUN} that, as in the SU(2) theory, the divergent part of the effective action reduces to its commutative counterpart.


\section{Analysing renormalisability}

In this section we will analyse whether the divergences in the effective actions, given in eqs.~\eqref{Gammatotaldiv} and \eqref{GammatotaldivSUN}, can be subtracted from appropriate multiplicative renormalisations of fields and parameters and infinite shifts on the Seiberg-Witten map ambiguities ${\mathfrak S,L}$ -see eq.~\eqref{SWmaps}. We will use the minimal subtraction scheme.  The counterterms in the action that cancel the divergences of the effective action are trivially given by
\begin{align*}
\iDx{\cal L}^{ct}=-\iDx\Gamma^{\rm div}.
\end{align*}
Were the theory to be renormalisable, these counterterms would arise from multiplicative renormalisation and from ambiguities of the SW maps, which, as will be argued, are equivalent to field redefinitions. We define the multiplicative renormalisation as
\begin{align}\label{Zs}
a_\m&= Z_a^{1/2}a_\m^{R}, & b_\m&=  Z_b^{1/2}b_\m^{R}, & \la&= Z_\la^{1/2}\la^{R}, & u&= Z_u^{1/2}u^{R}, &g&=Z_g g^R, & h&=Z_h h^R,
\end{align}
with $Z_i=1+\delta Z_i$. It is easily seen that gauge invariance forces
$$\delta Z_a=0.$$
On the other hand, the SW map ambiguities at order $h$, ${\mathfrak S}_\m$ and ${\mathfrak L}$, are given by terms transforming in the adjoint representation of the gauge group, with the appropriate mass dimensions and index structure, involving a contraction with $\th^{\a\b}$; under a $U(N)$ gauge transformation of the fields with gauge parameter ${\mathfrak c}=c^a T^a +\frac{\unit}{\sqrt{2N}}C,$ they transform as
\begin{align*}
s {\mathfrak S}_\m=i[c^aT^a,{\mathfrak S}_\m],\quad s {\mathfrak L}=i[c^aT^a,{\mathfrak L}].
\end{align*}
 We restrict ourselves to ambiguities that respect the parity transformation properties of the fields; doing so is justified since considering additional types of field redefinitions would yield terms in the modified action involving an odd number of $\epsilon^{\m\n\r\s}$ tensors, which do not appear in the divergent parts of the effective actions given in eqs.~\eqref{Gammatotaldiv} and \eqref{GammatotaldivSUN}  and thus need not  be considered when checking renormalisablity.
	The most general solution we found satisfying the specified requirements is of the form
\begin{align}\label{SL}
{\mathfrak S}_\m=S_\m+T_\m{\unit }, {\mathfrak L}=L+M {\unit },
\end{align}
such that
\begin{align}
\nonumber S_\m=&y_1\th^{\a\b}D_\m f_{\a\b}+y_2{\th_\m}^\a D^\n f_{\n\a}+y_3{\th_\m}^\a\{\bar\la_i,(\ga_\a\la)_i\}+iy_4\th^{\a\b}[\bar\la_i,(\ga_{\m\a\b}\la)_i]+iy_5{\th_\m}^\a\bar u\ga_\a \la\\
\nonumber &+y_6\th^{\r\s}\bar u\ga_{\m\r\s}\la,\,y_k\in\RR,\\
\nonumber\nonumber L=&k_1\th^{\a\b}\{f_{\a\b},\la\}+k_2\th^{\a\b}D^2\la+k_3\th^{\a\b}{\ga_\a}^\m[f_{\b\m},\la]+k_4\th^{\a\b}{\ga_\a}^\m\{f_{\b\m},\la\}+k_5\th^{\a\b}{\ga_\a}^\m\{D_\m,D_\b\}\la\\
\nonumber&+k_6\tilde{\theta}^{\a\b}\ga_5\{f_{\a\b},\la\}+k_7\tilde\theta^{\a\b}\ga_5[f_{\a\b},\la]
+k_8\th^{\a\b}g_{\a\b}\la+k_9\th^{\a\b}{\ga_\a}^\m g_{\b\m}\la+k_{10}\tilde\theta^{\a\b}\ga_5 g_{\a\b} \la\\
\label{SWambig}&+k_{11}\th^{\a\b}f_{\a\b}u+k_{12}\th^{\a\b}{\ga_\a}^\m f_{\b\m}u+k_{13}\tilde\theta^{\a\b}\ga_5f_{\a\b} u, \,k_i\in\mathbb{C},\\
\nonumber T_\m=&z_1\th^{\a\b}\Dm g_{\a\b}+z_2{\th_\m}^\a \partial^\n g_{\n\a}+i z_3\th^{\a\b}\bar u \ga_{\m\a\b}u+iz_4\Tr\th^{\a\b}[\bar\la_i,(\ga_{\m\a\b}\la)_i],\,z_k\in\RR,\\
\nonumber M=&l_1\th^{\a\b}g_{\a\b}u+l_2\th^{\a\b}\partial^2 u+l_3\th^{\a\b}{\ga_\a}^\m g_{\b\m}u+l_4\th^{\a\b}{\ga_\a}^\m\Dm\Db u+l_5\tilde\theta^{\a\b}\ga_5 g_{\a\b}u+l_6\Tr\th^{\a\b}\{f_{\a\b},\la\}\\
\nonumber &+l_7\Tr\th^{\a\b}{\ga_\a}^\m\{f_{\b\m},\la\}+l_8\Tr\tilde\th^{\a\b}\ga_5\{f_{\a\b},\la\},\,l_i\in\mathbb{C},
\end{align}
where $\tilde\theta^{\a\b}=\frac{1}{2}\epsilon^{\a\b\r\s}\th_{\r\s}$. In the U(N) case, since the enveloping algebra coincides with the Lie algebra, the previous ambiguities are equivalent to field redefinitions of $b_\m,a_\m,\la,u$.  The ambiguities in the SU(N) case are obtained by setting $b_\m=u=0$. Since in principle there still remain contributions along the identity operator after setting $b_\m=u=0$ in eq.~\eqref{SWambig}, it would seem that the SU(N) ambiguities are not equivalent to field redefinitions, which would invalidate our arguments concerning the possibility of setting the ambiguities to zero at tree-level without losing generality when dealing with the renormalisability of the theory. However, in the SU(N) case it is easily seen that the contributions to the ambiguities along the identity, coming from the terms of eq.~\eqref{SWambig} proportional to $y_4,k_1,k_4,k_6,z_4,l_6,l_7,l_8,$ do not yield modifications of the action at order $h$, so that these contributions can be ignored and the ambiguities can be thought as Lie algebra valued and thus equivalent to field redefinitions. Note that the field redefinitions proportional to $y_4,k_1,k_4,k_6$ not only yield contributions along the identity but also on the Lie Algebra, and thus still have to be taken into account.


\subsection{Commutative renormalisation}
	The U(N) and SU(N) --$N>1$-- theories at order $h=0$ are multiplicatively renormalisable; the divergences appearing in eq.~\eqref{Gammaord} -see also eq.~\eqref{GammaordSUN}- can be absorbed by the following values of the renormalisation constants in eq.~\eqref{Zs}:
\begin{align}\label{Zsol}
\delta Z_a&=0, &\delta Z_g&=-\frac{3g^2N}{32\pi^2\epsilon}, &\delta Z_b&=-\frac{3g^2N}{16\pi^2\epsilon}, &\delta Z_\la&=-\frac{g^2N}{4\pi^2\epsilon}, & \delta Z_u&=-\frac{3g^2N}{16\pi^2\epsilon}.
\end{align}


\subsection{Renormalisation of the noncommutative bosonic sector}
	Starting with the U(N) case, $N>1$, let us consider the order $h$ noncommutative divergences only involving gauge fields in eq.~\eqref{Gammatotaldiv}.  A key issue is that
the ambiguities in the SW map given in eq.~\eqref{SWambig}, when introduced in the action by means of eqs.~\eqref{SL}, \eqref{SWambig}, \eqref{SWmaps} and \eqref{Snc}, do not generate any purely bosonic terms.
    Thus the purely bosonic divergences can only be renormalised, if at all, by means of multiplicative renormalisations.
    In terms of the basis of gauge invariant terms given in eqs.~\eqref{ts}, \eqref{us} and \eqref{vs}, the tree-level noncommutative contribution to the bosonic part of the action is
 \begin{align}
\nonumber S_{\rm [U(N)] bos }^{\rm{tree},NC}=&\iDx\Big(\frac{h}{4g^2}t_1-\frac{h}{g^2}t_2+\frac{h}{8g^2\sqrt{2N}}u_1-\frac{h}{2g^2\sqrt{2N}}u_2+\frac{h}{4g^2\sqrt{2N}}(v_1+2v_4)\\
\label{UNbos}&-\frac{h}{g^2\sqrt{2N}}(v_2+2v_3)\Big)+O(h^2),
 \end{align}
so that the counterterm action originated by the multiplicative renormalisations of eq.~\eqref{Zs} would be, keeping in mind $\delta Z_a=0$ and suppressing the ``$R$" superindices for simplicity:
\begin{align*}
S_{\rm [U(N)] bos}^{\rm{ct},NC}=&\iDx\Big((-2 \delta Z_g+\delta Z_h)\Big(\frac{h}{4g^2}t_1-\frac{h}{g^2}t_2\Big)\\
&+\big(-2 \delta Z_g+\delta Z_h+\frac{3}{2}\delta Z_b\big)\Big(\frac{h}{8g^2\sqrt{2N}}u_1-\frac{h}{2g^2\sqrt{2N}}u_2\Big)\\
&+\big(-2 \delta Z_g+\delta Z_h+\frac{1}{2}\delta Z_b\big)\Big(\frac{h}{4g^2\sqrt{2N}}(v_1+2v_4)-\frac{h}{g^2\sqrt{2N}}(v_2+2v_3)\Big)\Big)+O(h^2),
\end{align*}
which should be made equivalent with minus the bosonic part of the divergent part of the effective action in eq.~\eqref{Gammadivts}
\begin{align}\nonumber
\Gamma^{div,NC}_{\rm[U(N)] bos}\!=\!&-\iDx\Big(\frac{3g^2N}{16\pi^2\epsilon}\Big[\frac{h}{4g^2}t_1-\!\frac{h}{g^2}t_2\Big]+\frac{2g^2N}{16\pi^2\epsilon}\Big[\frac{h}{4g^2\sqrt{2N}}(v_1+2v_4)-\frac{h}{g^2\sqrt{2N}}(v_2+2v_3)\Big]\\
&+O(h^2).\label{Gammadivbos}
\end{align}
For $N\geq3$, for which the terms $t_1$ and $t_2$ are nonzero, this forces the  three following identities,
\begin{align}
\nonumber \frac{3g^2N}{16\pi^2\epsilon}&=-2 \delta Z_g+\delta Z_h,\\
\label{gaugesectZ}0&=-2 \delta Z_g+\delta Z_h+\frac{3}{2}\delta Z_b,\\
\nonumber -\frac{2g^2N}{16\pi^2\epsilon}&=-2 \delta Z_g+\delta Z_h+\frac{1}{2}\delta Z_b.
\end{align}
Using the commutative results in eq.~\eqref{Zsol}, the first equation implies
$$\delta Z_h=0,$$
as has been obtained for a number of other noncommutative theories, but then the second and third identities in eq.~\eqref{gaugesectZ} are not satisfied. In the $N=2$ case, only the last two identities are relevant, since $t_1=t_2=0$; again, they are incompatible with the $O(h^0)$ results of eq.~\eqref{Zsol}.

	From this we arrive to the first conclusions of our paper: the U(N) theory is not renormalisable, for $N>1$.  In principle we have derived this only for our choice of gauge-fixing; to extend the result for arbitrary gauge-fixing, we have to consider the on-shell divergences, which are independent of the gauge-fixing. The equations of motion are of the form
\begin{align}\label{eom}
&(D_\m f^{\m\n})=\frac{1}{2} \{\bar\la_i,\ga^\n_{ij}\la_j\}+O(h),\quad \partial_\m g^{\m\n}=O(h),\quad \Dirac u=O(h),\\
\nonumber & (\Dirac\la)^a=\frac{h}{2}\th^{\a\b}\Tr T^a\g^\m\{D_\m\la,f_{\a\b}\}+h\th^{\a\b}\Tr T^a\ga^\m\{D_\b\la,f_{\m\a}\}+O(h^2),
\end{align}
where the details of the $O(h)$ part in the equations in the first line of \eqref{eom} will not be relevant to our purposes. Since the bosonic divergences of eq.~\eqref{Gammadivbos} --see eqs.~\eqref{ts} and \eqref{vs}-- do not involve covariant derivatives of field strenghts, it is easy to see that the on-shell conditions of eq.~\eqref{eom} cannot be used to relate the bosonic divergences in eq.~\eqref{Gammadivbos} among themselves or with the fermionic divergences. Thus the on-shell bosonic divergences have the same form as in eq.~\eqref{Gammadivbos} and the same conclusions about nonrenormalisability apply.

	  However, in the large N limit, $N\rightarrow\infty$ while keeping the 't Hooft coupling $g^2 N$ finite, both the tree-level contributions in eq.~\eqref{UNbos}  and the
problematic divergences  in eq.~\eqref{Gammadivbos} associated with the $u_i$ terms are subleading, so that the second and third identities in eq.~\eqref{gaugesectZ} need not be considered, and therefore the gauge sector is  renormalisable in this limit with $\delta Z_h=0,$ in keeping with the expectations raised by the quantum behaviour of the SW duals of the NC U(N) theories in the enveloping algebra approach.

	In the SU(N) case, $N\geq3$, we only have the divergences coming from the $t_i$ terms, which are multiplicatively renormalisable with $\delta Z_h=0$; thus, the SU(N) gauge sector is renormalisable, as has been obtained already for other NC theories in the enveloping algebra approach with different matter content \cite{Wulkenhaar:2001sq}, \cite{Buric:2002gm}, \cite{Buric:2004ms}, \cite{Buric:2005xe}, \cite{Buric:2006wm}, \cite{Martin:2006gw}.

	It remains to examine the renormalisability of the gaugino sector.  Given the previous conclusions,  it suffices to study only the large N limit of U(N) or the SU(N) theory for $N\geq3$, since the U(N) theory fails to be renormalisable for finite $N>1$.
	

\subsection{Renormalisation of the noncommutative gaugino sector}

In the large N limit, given the decomposition in eq.~\eqref{decomp}, the tree-level interactions involving U(1) fields are subleading with respect to those of SU(N) fields, so that at leading order in N the U(1) fields are free. This is also reflected at the quantum level, since the divergences involving U(1) fields in the effective action in eq.~\eqref{Gammatotaldiv}  are subleading. Thus, for large N the U(1) fields can be neglected, and the problem of renormalisability is the same as for the SU(N) theory.
The tree level part of the SU(N) action involving gaugino fields and taking into account the SW map ambiguities of eq.~\eqref{SWambig} restricted to the SU(N) case is given, for $N\geq3$ and in the basis of gauge invariant terms of eq.~\eqref{ts}, by
\begin{align*}
&S^{\rm tree,NC}_{\rm [SU(N)]gaugino}=\frac{h}{g^2}\iDx\sum_{i=3}^{16}C_i t_i+O(h^2),
\end{align*}\vskip-0.5cm
\begin{align*}
C_3&=-4i {\rm Re} k_{5}\!-\!4i{\rm Re} k_2, & C_4&=-2{\rm Im}k_2, \\
 C_5&=\frac{i}{2}\!-\!2i{\rm Re}k_4,&
C_6&=-\frac{i}{4}+2i{\rm Re}k_1, \\
 C_7&=-2i{\rm Re}k_4, & C_8&=-2iy_4+{\rm Re}k_6+i{\rm Im}k_6,\\
C_9&=\!-\frac{i}{2}{\rm Re}k_4+{\rm Re}k_6+i{\rm Im}k_6, & C_{10}&=\frac{1}{2}(2y_1+{\rm Im}k_3\!-{\rm Re}k_5), \\
C_{11}&=\frac{1}{2}(4y_3\!-2y_2+2{\rm  Im}k_3\!-\!2{\rm Re}k_5),&
C_{12}&=2{\rm Re}k_2+2i{\rm Im}k_2+2{\rm Re}k_6+2{\rm Re}k_7, \\
 C_{13}&=-2{\rm Re}k_5+2{\rm Re}k_6+2{\rm Re}k_7, & C_{14}&=2{\rm Im}k_3-2{\rm Re}k_5,\\
 C_{15}&=y_3, & C_{16}&=iy_{4}.
\end{align*}
The previous formulae follow from eq.~\eqref{Snc}, the SW map in eq.~\eqref{SWmaps} for ${\mathfrak a}_\m=a_\m$ and the ambiguities of eq.~\eqref{SWambig} for $T_\m=M=b_\m=u=0$. The Dirac algebra identities in eq.~\eqref{gammas} were extensively used.
	The counterterm Lagrangian generated by the multiplicative renormalisations of fields and parameters of eq.~\eqref{Zs} and by infinite shifts of the ambiguity parameters $y_i=y_i^R+\delta y_i^R, k_i=k_i^R+\delta k_i^R$ is given, for $y_i^R=k_i^R=0$ -recall that we computed the divergences in the effective action for zero values of the SW map ambiguities-, by
	\begin{align}\label{mattertree}
&S^{\rm ct,NC}_{\rm [SU(N)]gaugino}=\frac{h}{g^2}\iDx\sum_{i=3}^{16}(\delta C_i) t_i+O(h^2),
\end{align}%
\begin{align*}
\d C_3&=-4i \d{\rm Re} k_{5}-4i\d{\rm Re} k_2, & \d C_4&=-2\d{\rm Im}k_2,\\
 \d C_5&=\frac{i}{2}(-2\delta Z_g+\delta Z_\la) -2i\d{\rm Re}k_4, &
\d C_6&=-\frac{i}{4}(-2\d Z_g+\delta Z_\la)+2i\d {\rm Re}k_1,\\
\d C_7&=-2i\d{\rm Re}k_4, & \d C_8&=-2i\d y_4+\delta {\rm Re}k_6+i \delta{\rm Im}k_6,\\
\d C_9&=-\frac{i}{2}\d{\rm Re}k_4+\delta {\rm Re}k_6+i \delta{\rm Im}k_6, & \d C_{10}&=\frac{1}{2}(2\d y_1+\d{\rm Im}k_3-\d{\rm Re}k_5), \\
 \d C_{11}&=\frac{1}{2}(4\d y_3-2\d y_2+2\d{\rm  Im}k_3-2\d{\rm Re}k_5), & \d C_{12}&=2\d{\rm Re}k_2+2i\d{\rm Im}k_2+2\delta {\rm Re}k_6+2\delta{\rm Re}k_7,\\
 \d C_{13}&=-2\d{\rm Re}k_5+2\delta {\rm Re}k_6+2\delta{\rm Re}k_7, & \d C_{14}&=2\d{\rm Im}k_3-2\d{\rm Re}k_5,\\
 \d C_{15}&=\d y_3, & \d C_{16}&=i\d y_{4},
\end{align*}
where the ``$R$'' superindices have been suppressed for simplicity and the result $\delta Z_h=0$ of the previous subsection  was used.
	
	For the theory to be renormalisable, the previous counterterm action has to be matched with minus the divergent part of the effective action in eqs.~\eqref{GammadivtsSUN} and \eqref{GammatotaldivSUN} involving gaugino fields. This contribution involving fermion fields, expressed in the basis of gauge invariant terms of eq.~\eqref{ts}, is
\begin{align}\label{matterdiv}
\Gamma^{\rm div,NC}_{gaugino,[SU(N)]}=&\iDx\Big(\frac{iNh}{16\pi^2\epsilon}\Big[\frac{1}{4}t_6-\frac{1}{2}t_7-\frac{1}{8}t_8-\frac{1}{16}t_9\Big]+\frac{3iNh}{512\pi^2\epsilon}t_{16}\Big)+O(h^2).
\end{align}
Matching eqs.~\eqref{mattertree} and \eqref{matterdiv}, we get the following system of equations:
\begin{align}
\nonumber t_3&:\,\,\d {\rm Re}k_5+\d {\rm Re}k_2=0, & t_4&:\,\,\d {\rm Im}k_2=0, \\
\nonumber t_5&:\,\,\frac{i}{2}(-2\delta Z_g+\delta Z_\la) -2i\d{\rm Re}k_4=0, & t_6&:\,\,\frac{i}{4}(-2\d Z_g+\delta Z_\la)-2i\d {\rm Re}k_1=\frac{iNg^2}{64\pi^2\epsilon},\\
\nonumber t_7&:\,\,-2i\d{\rm Re}k_4=\frac{iNg^2}{32\pi^2\epsilon}, & t_8&:\,\,-2i\d y_4+\delta{\rm Re}k_6+i\delta{\rm Im}k_6=\frac{iNg^2}{128\pi^2\epsilon},\\
\nonumber  t_9&:\,\,-\frac{i}{2}\d{\rm Re}k_4+\delta{\rm Re}k_6+i\delta{\rm Im}k_6=\frac{iNg^2}{256\pi^2\epsilon}, & t_{10}&:\,\,2\d y_1+\d{\rm Im}k_3-\d{\rm Re}k_5=0,\\
\nonumber t_{11}&:\,\,4\d y_3-2\d y_2+2\d{\rm  Im}k_3-2\d{\rm Re}k_5=0, & t_{12}&:\d{\rm Re}k_2+i\d{\rm Im}k_2+\delta{\rm Re}k_6+\delta{\rm Re}k_7=0,\\
 \nonumber t_{13}&:\,\,-\d{\rm Re}k_5+\delta{\rm Re}k_6+\delta{\rm Re}k_7=0, & t_{14}&:\,\,\d{\rm Im}k_3-\d{\rm Re}k_5=0,\\
\label{tdivergences} t_{15}&:\,\,\delta y_3=0, &t_{16}&:i\d y_{4}=-\frac{3iNg^2}{512\pi^2\epsilon}.
\end{align}

Projecting into real and imaginary parts, there are 17 real equations for 13 real variables, $$\delta y_1, \delta y_2,\delta y_3,\delta  y_4,\delta{\rm Re}k_1,\delta{\rm Re}k_2,\delta{\rm Im}k_2,\delta{\rm Im}k_3,\delta{\rm Re}k_4,\delta{\rm Re}k_5,\delta{\rm Re}k_6,\delta{\rm Im}k_6,\delta{\rm Re}k_7.$$
\indent Remarkably, the equations are not independent and  there is a one-paramter family of solutions,
\begin{align*}
\delta y_1&=\delta y_2=\delta y_3=\delta{\rm Im}k_2=\delta{\rm Re}k_6=0,\\
\delta y_4&=-\frac{3g^2N}{512\pi^2\epsilon},\\
\delta {\rm Im}k_3&=\delta {\rm Re}k_7=-\delta{\rm Re}k_2=\delta{\rm Re}k_5,\\
\delta{\rm Re}k_1&=\delta{\rm Re}k_4=-\frac{g^2N}{64\pi^2\epsilon},\\
\delta{\rm Im}k_6&=-\frac{Ng^2}{256\pi^2\epsilon}.
\end{align*}

This shows that, at one-loop and order $\theta$, the U(N) theory in the large N limit and the SU(N) theory are renormalisable.



\section{Conclusions}

	In this paper we have calculated the $O(\th)$ divergent part of the background field effective action for the classical dual under the Seiberg-Witten map of  noncommutative ${\cal N}=1$ U(N) super Yang-Mills, as well as for the SU(N) theory that results from suppressing the U(1) degrees of freedom in the former U(N) theory. Our results
can be summarised as follows:  the quantisation of the classical dual under the Seiberg-Witten map of  ${\cal N}=1$
U(N), $\rm N>1$,  super Yang-Mills yields an  $U(N)$ supersymmetric ordinary quantum theory that is not renormalisable --and neither is
its gauge sector-- for finite values of $N$. In the large N limit, however, the U(N) theory remarkably becomes renormalisable. On the other hand, the SU(N) theory for arbitrary $N>2$ also turns out to be renormalisable.

	That both  --i.e., the standard, quantised in terms of noncommuative fields, and the $\theta$-expanded, defined by means of the Seiberg-Witten map-- large N supersymmetric U(N) theories are one-loop renormalisable and have the same running coupling constant hints at the fact that the classical  duality between noncommutative theories established by the Seiberg-Witten map may survive at the quantum level: as was mentioned in the introduction, the noncommutative U(N) super Yang-Mills theory, at one-loop and large N, is renormalisable and has a smooth $\theta^{\mu\nu}\rightarrow 0$ limit, and this behaviour is reproduced in the dual theory formulated by means of the Seiberg-Witten map. This is in agreement with previous studies regarding the survival at the quantum level of the Seiberg-Witten map duality, although they focused on UV finite theories such as noncommutative Chern-Simons \cite{Kaminsky:2003qq}.
	
	On the other hand, the SU(N) result represents the first case in the literature in which a noncommutative theory in the enveloping algebra approach involving fermion fields turns out to be one-loop renormalisable at order $\theta$. This could be attributed to the consideration of Majorana fermions and a noncommutative covariant derivative involving a star product commutator, or perhaps, more likely, to the  effects  surviving in the large N limit of the supersymmetry present in the parent U(N) theory. Though the role of supersymmetry in the SU(N) case asks for futher analysis, the fact that the fermionic divergences can be renormalised, in contrast to the cases previously studied in the literature --in which the bosonic sector was found to be renormalisable, but not so the fermionic sector-- suggests that a symmetry relating fermions and bosons may actually be present, making the corresponding divergences not independent.  Our result encourages the study of noncommutative models sharing features with the SU(N) theories studied here, with either Majorana fermions or supersymmetry in terms of noncommutative fields; it raises the hope of constructing renormalisable noncommutative  gauge theories in the enveloping algebra approach with matter fields. Also, the effect on the ordinary fields of  the supersymmetry of the noncommutative fields in the SU(N) case is still not understood and needs further investigation.


\section{Acknowledgements}

This work has been financially supported in part by MICINN through grant
FPA2008-04906, and by the National Science Foundation under Grant No. PHY05-51164. The work of C. Tamarit has also received financial support from MICINN and the Fulbright Program through grant 2008-0800.
\newpage
\appendix
\section{Lie algebra, Dirac algebra, Majorana spinors}
 Following the notation of ref.~\cite{Bonora:2000ga}, we denote the Lie algebra generators of U(N) in the fundamental representation as $T^A=\{T^0,T^a\}, a=1,\dots,N^2-1,$ with $T^0=\frac{\unit}{\sqrt{2N}}$ and $T^a$ the standard SU(N) generators. The generators satisfy
\begin{align*}
[T^A,T^B]=i f^{ABC}T^C,\quad\{T^A,T^B\}=d^{ABC}T^C,
\end{align*}
where $f^{ABC}$ are totally antisymmetric, with $f^{abc}$ having their usual SU(N) values and $f^{0BC}=0,$ whereas $d^{ABC}$ are totally symmetric,  $d^{abc}$ having their usual SU(N) values and $d^{0BC}=\sqrt{2/N}\delta^{BC},\,d^{00c}=0,\,d^{000}=\sqrt{2/N}.$ We will make use of the following identities:
\begin{align}
\nonumber &f^{ACD}f^{BCD}=Nc_A\delta^{AB},\\
\label{UNcont} & f^{DAE}f^{EBF}f^{FCD}=-\frac{N}{2}f^{ABC},\\
  \nonumber&f^{DAE}f^{EBF}d^{FCD}=-\frac{N}{2}d^{ABC}c_Ac_Bd_C,\\
\nonumber&c_A=1-\delta_{A,0},\,\, d_A=2-c_A.
\end{align}

	Concerning Dirac $\ga$ matrices, satisfying $\{\ga^\m,\ga^n\}=2\eta^{\m\n},$ we use a basis of operators in the space of spinors constructed from antisymmetrised products of these matrices: $\{\ga^\m, \ga^{\m\n},\ga^{\m\n\r},\ga_5\},$ with the following definitions
\begin{align*}
\ga^{\m\n}&=\frac{1}{2}(\ga^\m\ga^\n-\ga^\n\ga^\m),& \!\!\!\!\ga^{\m\n\r}&=\frac{1}{6}(\ga^\m\ga^\n\ga^\r+\ga^\n\ga^\r\ga^\m+\ga^\r\ga^\m\ga^\n-\ga^\m\ga^\r\ga^\n-\ga^\n\ga^\m\ga^\r-\ga^\r\ga^\n\ga^\m),\\
\ga_5&=-\frac{i}{4!}\epsilon_{\m\n\r\s}\ga^\m\ga^\n\ga^\r\ga^\s.
\end{align*}
	In order to express products of $\ga$ matrices in terms of the previous basis, the following identities can be used:
\begin{align}
\nonumber&\ga^\m\ga^\n=\eta^{\m\n}+\ga^{\m\n},\\
\nonumber&\ga^\m\ga^\n\ga^\r=\eta^{\n\r}\ga^\m-\eta^{\m\r}\ga^\n+\eta^{\m\n}\ga^\r+\ga^{\m\n\r},\\
\nonumber&\ga^\m\ga^\n\ga^\la\ga^\r=\eta^{\m\n}\eta^{\la\r}+\eta^{\m\r}\eta^{\n\la}-\eta^{\m\la}\eta^{\n\r}+\eta^{\m\n}\ga^{\la\r}
+\eta^{\m\r}\ga^{\n\la}-\eta^{\m\la}\ga^{\n\r}+\eta^{\n\la}\ga^{\m\r}-\eta^{\n\r}\ga^{\m\la}+\eta^{\la\r}\ga^{\m\n}\\
\label{gammas}&\phantom{\ga^\m\ga^\n\ga^\r\ga^\la=}-i\epsilon^{\m\n\la\r}\ga_5.
\end{align}
	Majorana spinors are self-conjugate, satisfying
\begin{align}
\label{Maj1}
\la=C\bar\la^T,\quad\bar\la=-\la^T C^{-1}
\end{align}
for a charge conjugation matrix $C$ such that
\begin{align}
C^\dagger=C^{-1},\quad C^T=-C,\quad C\Gamma_i^T C^{-1}=\eta_i \Gamma_i,\,\,\eta_i=\left\{\begin{array}{ll}
+1,\,\Gamma_i={\rm{I}\!\rm{I}},\ga_5,\ga^{\m\n\r}\\
-1,\,\Gamma_i=\ga^{\m},\ga^{\m\n}.
\end{array}\right.\label{Maj2}
\end{align}
\section{Feynman rules for $\alpha=1,$ ${\mathfrak S}_{\m}^{\rm tree}= {\mathfrak L}^{\rm tree}=0$.}	
The background field legs are denoted by an encircled "b". We define the Feynman rules without symmetrising over these background field legs, which is consistent with the definition of the expansion of the effective action in terms of diagrams provided in eq.~\eqref{Gammaexp}. Since we are doing one-loop calculations, only vertices with two quantum gauge  fields contribute; vertices with one quantum gauge field are ignored since they do not contribute to 1PI diagrams. Since we are dealing with self-conjugate Majorana fermions, the vertices with Majorana fermions have to be symmetrised with respect to the conjugation of the interaction in each fermion pair, using \eqref{Maj2} \cite{Denner:1992me}. The Feynman rules used in our computations are then the following:
\psfrag{p}{$p$}
\psfrag{q}{$q$}
\psfrag{k}{$k$}
\psfrag{k1}{$k_1$}
\psfrag{k2}{$k_2$}
\psfrag{k3}{$k_3$}
\psfrag{m A}{$\mu, A$}
\psfrag{m C}{$\mu, C$}
\psfrag{n B}{$\nu, B$}
\psfrag{l C}{$\la, C$}
\psfrag{r C}{$\r, C$}
\psfrag{r D}{$\r, D$}
\psfrag{i A}{\hskip-0.3cm$i, A$}
\psfrag{j B}{$j, B$}
\psfrag{A}{$A$}
\psfrag{B}{$B$}

\begin{minipage}{0.27\textwidth}
\includegraphics[scale=1]{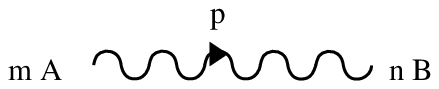}
\end{minipage}%
\begin{minipage}{0.23\textwidth}
\begin{align*}
\leftrightarrow\frac{-ig^2 \delta^{AB}\eta^{\m\n}}{p^2+i\epsilon}
\end{align*}
\end{minipage}%
\begin{minipage}{0.3\textwidth}
\includegraphics[scale=0.7]{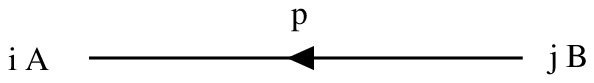}
\end{minipage}%
\begin{minipage}{0.16\textwidth}
\begin{align*}
\leftrightarrow\frac{ig^2 (\pslash)_{ij}\delta^{AB}}{p^2+i\epsilon}
\end{align*}
\end{minipage}\\
\begin{minipage}{0.34\textwidth}
\hskip0.5cm\includegraphics[scale=0.8]{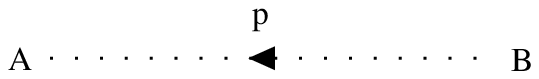}
\end{minipage}%
\begin{minipage}{0.2\textwidth}
\begin{align*}
\leftrightarrow\frac{i \delta^{AB}}{p^2+i\epsilon}
\end{align*}
\end{minipage}\\
\begin{minipage}{0.4\textwidth}
\flushleft\hskip0.5cm
\includegraphics[scale=0.7]{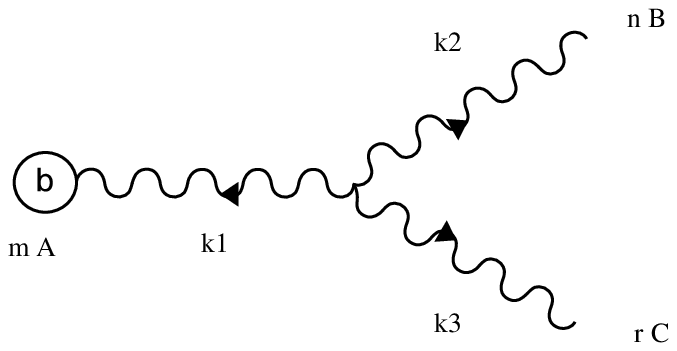}
\end{minipage}%
\begin{minipage}{0.55\textwidth}\flushleft
\begin{align*}
\leftrightarrow&\frac{1}{g^2}f^{ABC}[\eta^{\m\r}(k_1-k_3-k_2)^\n+\eta^{\n\r}(k_3-k_2)^\m\\
&+\eta^{\m\n}(k_2-k_1+k_3)^\r]
\end{align*}
\end{minipage}\\
\vskip0.2cm
\begin{minipage}{0.4\textwidth}
\flushleft
\includegraphics[scale=0.55]{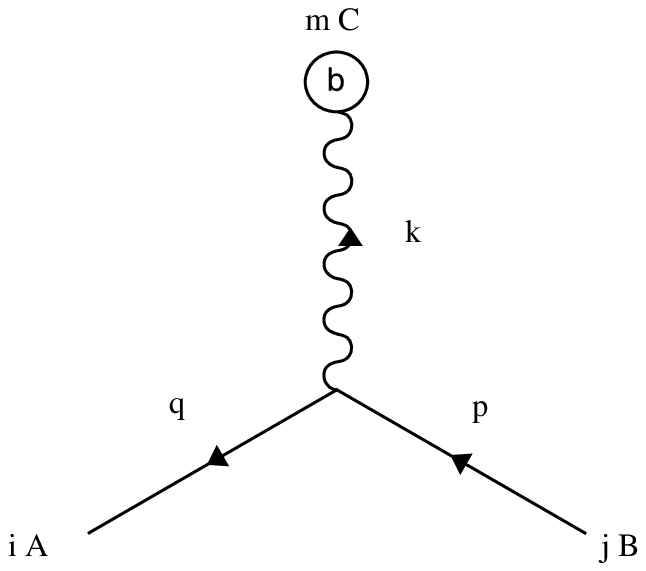}
\end{minipage}%
\begin{minipage}{0.56\textwidth}
\begin{align*}
&\leftrightarrow\frac{1}{g^2}(\ga^\m)_{ij}f^{ABC}
\end{align*}
\end{minipage}
\vskip0.2cm
\begin{minipage}{0.4\textwidth}
\flushleft\hskip-0.2cm
\includegraphics[scale=0.7]{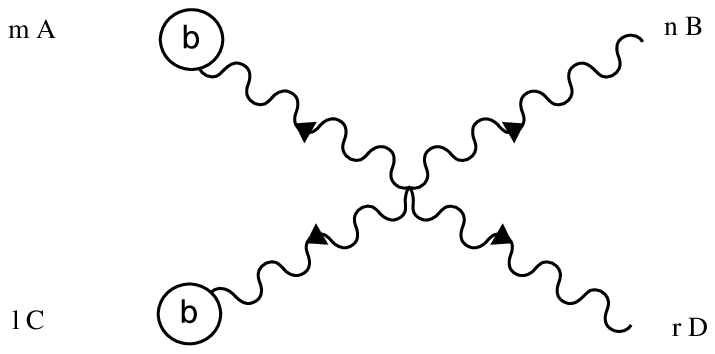}
\end{minipage}%
\begin{minipage}{0.56\textwidth}\flushleft
\begin{align*}
\leftrightarrow-\frac{i}{2g^2}[&f^{ABF}f^{FCD}(\eta^{\m\la}\eta^{\n\r}-\eta^{\m\r}\eta^{\n\la}+\eta^{\m\n}\eta^{\la\r})\\
&f^{ADF}f^{FBC}(\eta^{\m\n}\eta^{\la\r}-\eta^{\m\la}\eta^{\n\r}-\eta^{\m\r}\eta^{\n\la})]\\
&f^{ACF}f^{FBD}(\eta^{\m\n}\eta^{\la\r}-\eta^{\m\r}\eta^{\n\la})]
\end{align*}
\end{minipage}\\\vskip0.1cm
\begin{minipage}{0.33\textwidth}
\flushleft\hskip0.1cm
\includegraphics[scale=0.7]{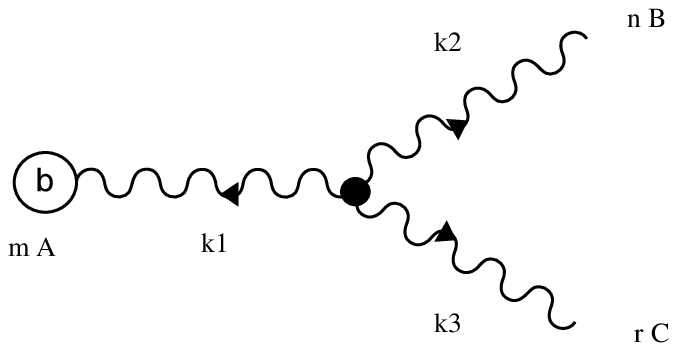}
\end{minipage}%
\begin{minipage}{0.63\textwidth}
\begin{align}
\nonumber\leftrightarrow&\frac{1}{4g^2}\th_{\a\b}d^{ABC}[k_1^{\a}k_2\cdot k_3\eta^{\m\b}\eta^{\n\r}-k_1^\a k_2^\r k_3^\n \eta^{\m\b}\\
\label{3gaugefr}&-2(k_1^\a k_2^\b k_3^\m \eta^{\n\r}-k_1^\a k_2^\r k_3^\m \eta^{\n\b}-k_1\cdot k_3 k_2^\b \eta^{\n\r}\eta^{\m\a}\\
\nonumber&+k_1\cdot k_3 k_2^\r \eta^{\m\a}\eta^{\n\b})]+(\text{permutations of all legs})
\end{align}
\end{minipage}\\
\begin{minipage}{0.35\textwidth}
\flushleft\hskip0.6cm
\includegraphics[scale=0.6]{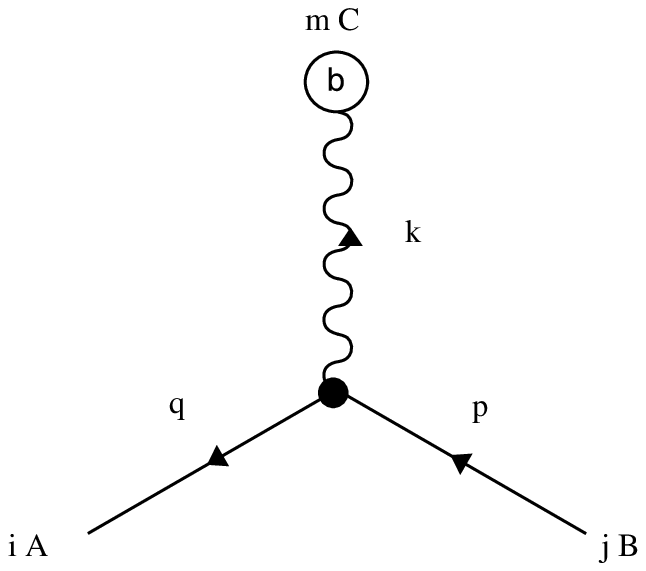}
\end{minipage}%
\begin{minipage}{0.61\textwidth}
\begin{align*}
\leftrightarrow&\frac{1}{4g^2}\th_{\a\b}d^{ABC}(\ga_\r)_{ij}[-\eta^{\m\a}q^\r p^\b+2\eta^{\r\m} q_\a p_\b+\eta^{\m\a}p_\r q_\b\\
&\phantom{\frac{1}{4g^2}\th_{\a\b}d^{ABC}(\ga_\r)_{ij}[}+\eta^{\m\a}p^\r q^\b-\eta^{\m\a}q_\r p_\b]
\end{align*}
\end{minipage}\\
\begin{minipage}{0.4\textwidth}
\flushleft\hskip-0.2cm
\includegraphics[scale=0.7]{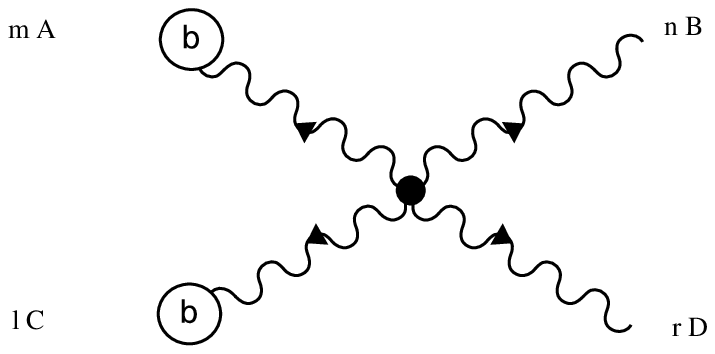}
\end{minipage}%
\begin{minipage}{0.56\textwidth}
\begin{align*}
\leftrightarrow&\frac{-i}{16g^2}\th_{\a\b}f^{ABF}d^{CDF}[k_3\cdot k_4\eta^{\m\a}\eta^{\n\b}\eta^{\la\r}-k_3^\r k_4^\la \eta^{\m\a}\eta^{\n\b}\\
&+4k_3^\a k_4^\m \eta^{\n\r}\eta^{\la\b}-4(k_3^\b k_4 ^\n \eta^{\m\a}\eta^{\la\r}-k_3^\b k_4 ^\la \eta^{\m\a}\eta^{\n\r}\\
&-k_3^\r k_4 ^\n \eta^{\m\a}\eta^{\la\b}+k_3\cdot k_4\eta^{\m\a}\eta^{\n\r}\eta^{\la\b})-2(k_3^\a k_4^\b \eta^{\m\la}\eta^{\n\r}\\
&-k_3^\a k_4^\n \eta^{\m\la}\eta^{\r\b}-k_3^\m k_4^\b \eta^{\n\r}\eta^{\la\a}+k_3^\m k_4^\n \eta^{\la\a}\eta^{\r\b})]\\
&+(\text{permutations of all legs})
\end{align*}
\end{minipage}\\\vskip0.1cm
\begin{minipage}{0.82\textwidth}
\flushleft\hskip-0.2cm\psfrag{m A}{$\m, C$}\psfrag{l C}{$\n, D$}
\includegraphics[scale=0.7]{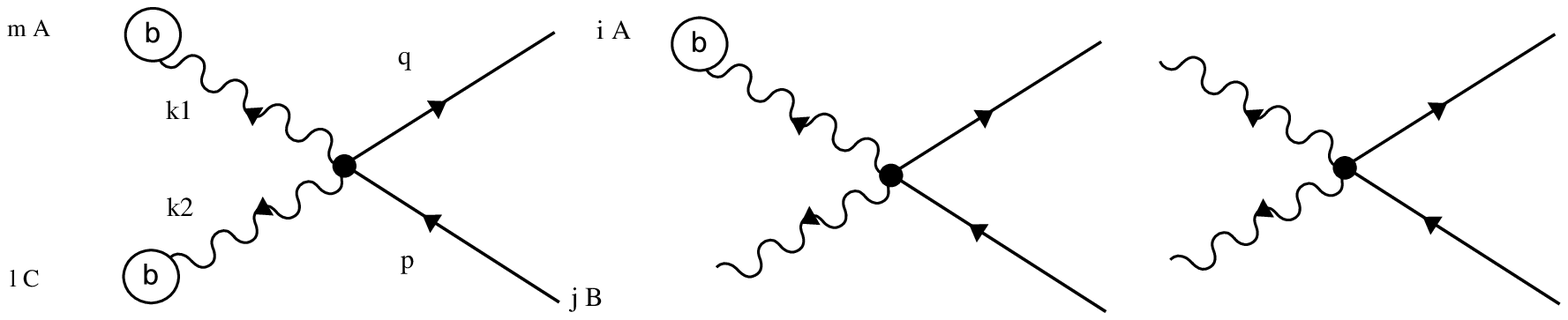}
\end{minipage}%
\begin{minipage}{0.12\textwidth}$\leftrightarrow$
\end{minipage}\\
\begin{minipage}{0.98\textwidth}
\begin{align*}
\leftrightarrow&\frac{-i}{2g^2}\th_{\a\b}(\ga_\r)_{ij}\Big[\frac{1}{2}(d^{ACE}f^{BDE}-d^{BCE}f^{ADE})[k_1^\a(\eta^{\m\b}\eta^{\r\n}-\eta^{\r\m}\eta^{\n\b})+k_1^\r\eta^{\m\a}\eta^{\n\b}]\\
&+\frac{1}{2}(d^{ADE}f^{BCE}-d^{BDE}f^{ACE})[k_2^\a(\eta^{\n\b}\eta^{\r\m}-\eta^{\r\n}\eta^{\m\b})+k_2^\r\eta^{\n\a}\eta^{\m\b}]\\
&+\frac{1}{2}d^{ABE}f^{CDE}[(p+q)^\r\eta^{\m\a}\eta^{\n\b}+(p+q)^\a(\eta^{\m\b}\eta^{\r\n}-\eta^{\n\b}\eta^{\r\m})]\Big]
\end{align*}
\end{minipage}

\end{document}